\documentclass[11pt]{article}
\pdfoutput=1
\usepackage{hyperref}
\usepackage{float}
\usepackage{alphalph}
\usepackage[utf8]{inputenc}
\usepackage{graphicx}
\usepackage{amssymb}
\usepackage{amsmath}
\usepackage{booktabs}
\usepackage{multirow}
\usepackage[title]{appendix}
\usepackage{dcolumn}
\newcolumntype{d}[1]{D{.}{.}{#1}}

\usepackage{soul,color} 
\usepackage{xcolor}
\sethlcolor{yellow} 
\usepackage{transparent}
\begin{document}
\pagenumbering{arabic}
\centerline{\LARGE EUROPEAN ORGANIZATION FOR NUCLEAR RESEARCH}
\vspace{15mm}
 \begin{flushright}
 CERN-EP-2023-032\\
March 16, 2023\\
 \vspace{2mm}
Revised version:\\
 \today \\
  \vspace{2mm}
 \end{flushright}
\vspace{15mm}

\begin{center}
\Large{\bf Search for dark photon decays to $\mathbf{\mu^+\mu^-}$ at NA62 \\
\vspace{5mm}
}
The NA62 Collaboration\renewcommand{\thefootnote}{\fnsymbol{footnote}} %
\footnotemark[1]\renewcommand{\thefootnote}{\arabic{footnote}}
\end{center}
\vspace{10mm}

\begin{abstract}
The NA62 experiment at CERN, designed to study the ultra-rare decay $K^{+} \rightarrow \pi^{+}\nu\bar{\nu}$, has also collected data in beam-dump mode. 
In this configuration, dark photons
may be produced by protons dumped on an absorber and reach a decay volume
beginning 80~m downstream. A search for dark photons decaying in flight to $\mu^+\mu^-$ pairs is reported, based on a sample of $1.4\times 10^{17}$ protons on dump collected in 2021. 
No evidence for a dark photon signal is observed. A
region of the parameter space is excluded at 90\% CL, improving on
previous experimental limits for dark photon masses between 215 and 550~MeV$/c^2$.
\end{abstract}
\vspace{40mm}

\renewcommand{\thefootnote}{\fnsymbol{footnote}}
\footnotetext[1]{email: na62eb@cern.ch \\ Corresponding authors: B.~D\"obrich, E.~Minucci, T.~Spadaro, email:  babette.dobrich@cern.ch, elisa.minucci@cern.ch, tommaso.spadaro@cern.ch}
\renewcommand{\thefootnote}{\arabic{footnote}}

\clearpage
\section{Introduction}
\label{sec:intro}
Proposed extensions of the Standard Model (SM) aimed at explaining the abundance of dark matter in the universe predict an additional $U(1)$ gauge-symmetry sector with a vector mediator field $A^\prime$, often called a ``dark photon".
In a simple realization of such a scenario~\cite{Okun,Holdom}, the $A^\prime_{\mu}$ field with mass $M_{A^\prime}$ interacts with the gauge field $B^{\mu}$ associated with the SM $U(1)$ symmetry through a kinetic-mixing Lagrangian term:
\begin{equation}
   -\frac{\varepsilon}{2} \, F^{\prime}_{\mu\nu}{B}^{\mu\nu},
\end{equation}
where $F^{\prime}_{\mu\nu} = \partial_{\mu}A^{\prime}_{\nu}-\partial_{\nu}A^{\prime}_{\mu}$, $B^{\mu\nu} = \partial^{\mu}B^{\nu}-\partial^{\nu}B^{\mu}$, 
and $\varepsilon\ll 1$ is the coupling constant. The mass $M_{A^\prime}$ and the coupling constant $\varepsilon$ are the free parameters of the model. 
The relevant features of the dark photon phenomenology are:
\begin{itemize}
    \item Dark photons can be produced in proton-nucleus interactions via bremsstrahlung or decays of secondary mesons. The two mechanisms differ in terms of production cross-section and distributions of the momenta and angles of the dark photons. At the energy of SPS protons (400~GeV), the probability for production of a dark photon with a momentum above 10~GeV$/c$ is of the order of $10^{-2}\times\varepsilon^2$ per proton.
    \item For $\varepsilon$ in the range from $10^{-7}$ to $10^{-5}$ and $M_{A^\prime}$ in the range from MeV$/c^2$ to GeV$/c^2$, the decay lengths of dark photons with momenta above 10~GeV$/c$
    span from tens of metres to tens of kilometres.
    \item Due to the feeble interaction with SM particles, dark photons can punch through tens of metres of material before decaying.
    \item For $M_{A^\prime}$ below 700~MeV$/c^2$, the dark photon decay width is dominated by di-lepton final states.
\end{itemize}
Other new-physics scenarios can lead to di-lepton final states.
Proton beam-dump experiments are a high-intensity source of secondary muons, providing an opportunity to probe muon-specific dark sectors~\cite{Rella:2022len}. 
Another scenario, which is considered here, is the proton-induced emission of axion-like particles (ALP) coupled to SM fermionic fields~\cite{Dobrich:2018jyi}.
An ALP $a$ can be emitted in the decays of charged or neutral $B$ mesons produced in proton-nucleus interactions:
\begin{equation}
\label{eq:alpProd}
    p N \to B X, ~~\mbox{followed by }B\to K^{(\ast)} a.
\end{equation}
ALPs with masses below 700~MeV$/c^2$ and interacting only with SM fermionic fields decay mainly to di-lepton modes. To address the general scenario in which the coupling of ALPs to SM fermionic fields is not uniform (for example, the coupling to quarks differs from that to leptons), a model-independent approach is adopted: the product of branching ratios 
\begin{equation}
\label{eq:alpBRproduct}
    \mathrm{BR}(B\to K^{(\ast)} a)\times\mathrm{BR}(a\to\mu^+\mu^-)
\end{equation}
is assumed to be independent of the $a$ lifetime. The free parameters in this case are the $a$ mass and lifetime, and the product of the branching ratios of eq.~(\ref{eq:alpBRproduct}).\\

The intense proton beam extracted from the CERN SPS and the NA62 setup have been exploited to search for the production and decay of dark photons by taking data in beam-dump mode: 
 $1.4\times10^{17}$ protons were dumped in 10 days in 2021.
The first NA62 search for dark photon decays to di-muon final states in beam-dump mode is presented. 

\section{Beam-dump operation of NA62}
\begin{figure}[t]
\begin{center}
\includegraphics[width=.95\textwidth]{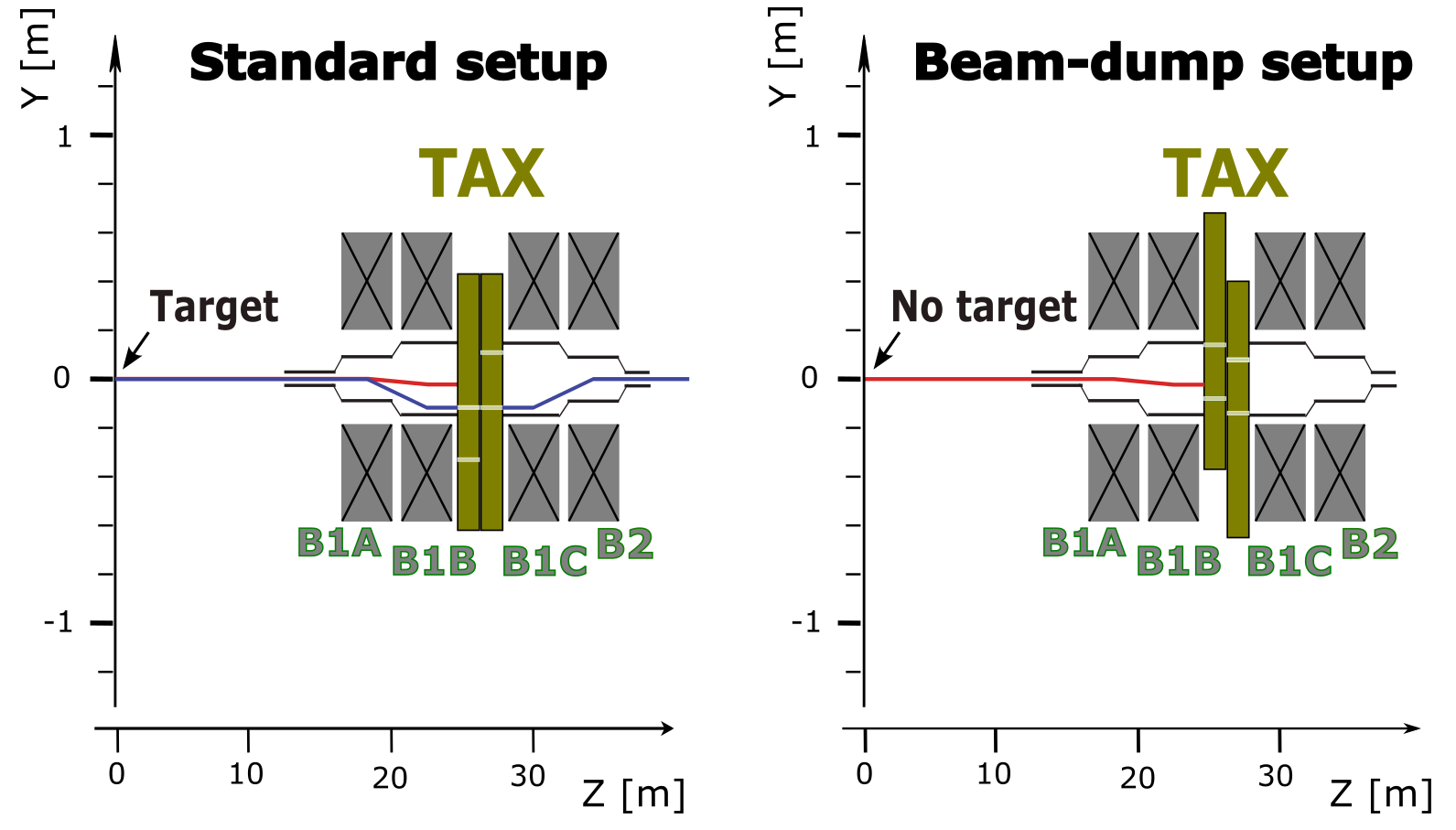}
\caption{\label{fig:TAXRegion} Schematic Y-Z view of the TAX achromat: standard (left) and beam-dump (right) setups. The holes in the TAX movable parts are aligned (left) and misaligned (right). The beam enters from the left. The trajectory of a proton with 400~GeV$/c$ momentum along Z at the origin is drawn in red. In the left panel, the trajectory of a particle with positive charge and 75 GeV$/c$ momentum along Z at the origin is drawn in blue.
}
\end{center}
\end{figure}

In the standard operation, dedicated to the study of the $K^{+} \rightarrow \pi^{+}\nu\bar{\nu}$ decay, a 400~GeV proton beam extracted from the CERN SPS is focused onto a 400~mm long, 2~mm diameter beryllium rod to generate a secondary hadron beam. An achromat composed of two movable collimators (TAX) located  between two pairs of dipoles is used for momentum selection, as sketched in the left panel of figure~\ref{fig:TAXRegion}. The origin of the coordinate system is at the target centre, the Z axis is directed downstream along the beam line, the Y axis points upwards, the X-Y-Z axes form a right-handed coordinate system. The dipoles upstream of the TAX (B1A, B1B) 
produce a downward translation of the beam axis, with a vertical shift inversely proportional to the particle momentum. The TAX holes are used to select beam particles in a narrow momentum range centred at 75~GeV$/c$. The dipoles downstream of the TAX (B1C, B2) shift the beam back to the original axis.  

\begin{figure}[t]
\begin{center}
\includegraphics[width=.7\textwidth]{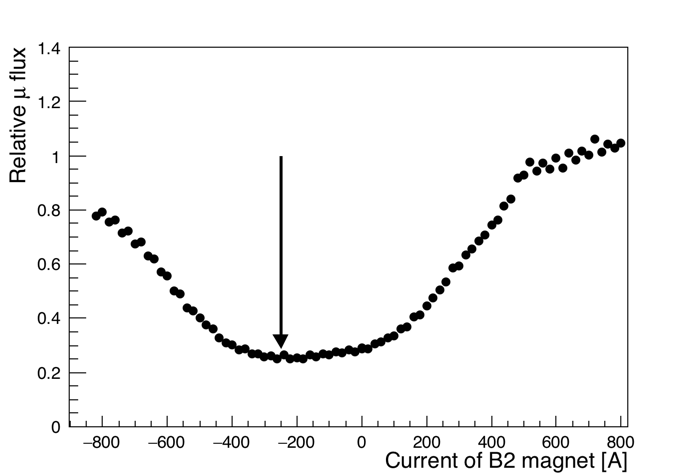}
\caption{\label{fig:MuonFlux} Relative muon flux measured by the MUV3 detector (section~\ref{sec:bl}) as a function of the B2 magnet current. The reference point for the standard operation is $+770$~A, corresponding to a field strength of 1.8~T. The arrow indicates the working point for the beam-dump data taking, $-250$~A.}
\end{center}
\end{figure}

In the beam-dump operation, sketched in the right panel of figure~\ref{fig:TAXRegion}, the target is removed and
the holes in the two movable sections of the TAX are misaligned with respect to each other and the beam axis. The proton beam is dumped on 800~mm of copper followed by 2400~mm of iron, corresponding to a total of 19.6 nuclear interaction lengths.
The currents of the dipoles preceding the TAX are set as in the standard operation. 
The coordinates of the average proton impact point at the TAX front plane are
\begin{equation}
\label{eq:beamImpact}
P_0 = (0,-22,23070)~\mathrm{mm},
\end{equation}
with standard deviations of 4.7 and 3.2~mm in X and Y, respectively~\cite{PBCConventionalBeam}. The beam axis at the impact point is parallel to the Z axis.
In the beam-dump operation (unlike in the standard mode) the currents of the B1C and B2 dipoles are set to produce magnetic fields in the same direction. The magnetic field strength generated by B1C (B2) is 
$-1.8$~T ($-0.6$~T) along X to minimise the flux of ``halo'' muons produced by pion decays within the TAX, as predicted by simulations~\cite{Rosental}. The measurement of the muon flux relative to the standard operation as a function of the B2 current has confirmed the prediction (figure~\ref{fig:MuonFlux}). 

\subsection{NA62 beam line and detector}
\label{sec:bl}
\begin{figure}[t]
\includegraphics[width=.99\textwidth]{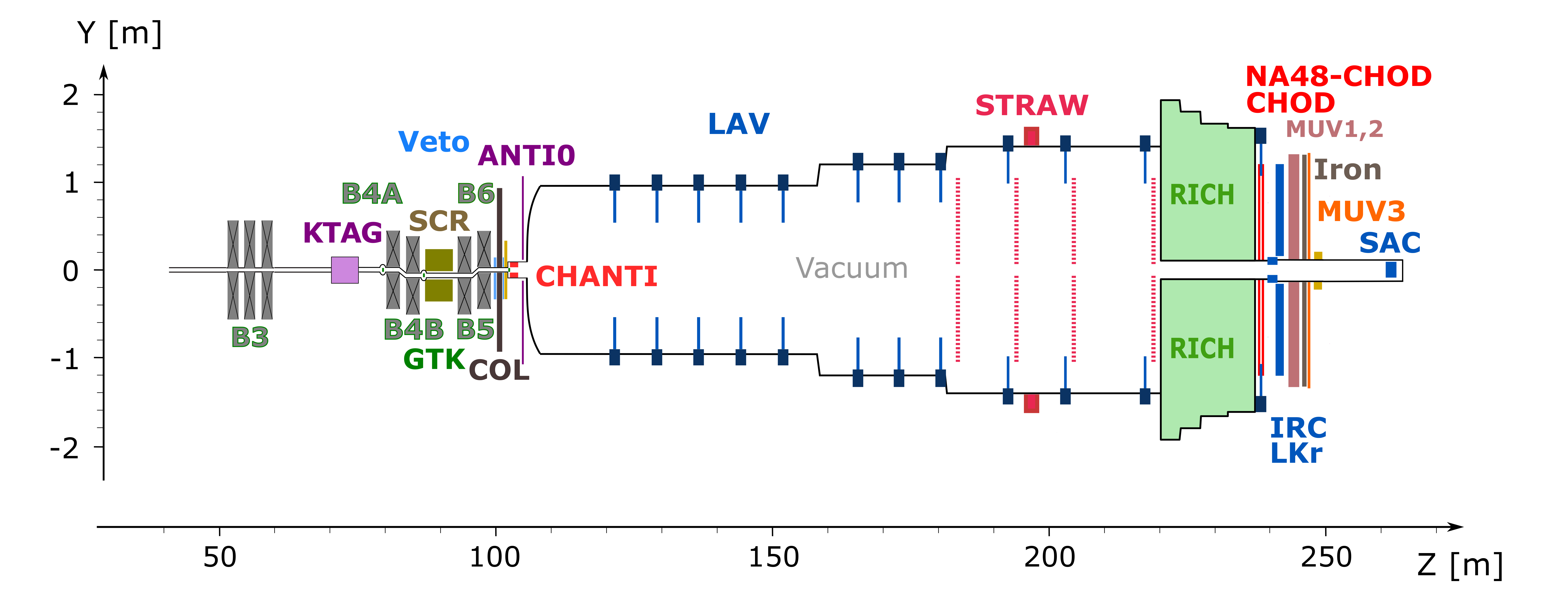}
\caption{\label{fig:layoutyz21} Schematic layout in the Y-Z plane of the NA62 experiment for the 2021 data taking. Certain elements of the beam line are not shown.}
\end{figure}

The beam line and detector~\cite{na62det} are sketched in figure~\ref{fig:layoutyz21}.
The elements relevant for the beam-dump operation are discussed here.

In addition to the dipole pair B1C-B2, other elements increase the capability of the beam line to sweep halo muons away from the detector acceptance. The elements with the highest sweeping power are: 
a triplet of magnetization-saturated dipole magnets (B3);  
 a toroidally-magnetized iron collimator (SCR) and the return yokes of the B5 and B6 magnets in the beam-tracker region (GTK, not used for this analysis). The cleaning collimator preceding the most downstream GTK station (COL, a 1.2~m thick steel block with outer dimensions $1.7\times1.8$~m$^2$) and the newly-installed ANTI0 scintillator hodoscope~\cite{ANTI0} are used 
to intercept and detect  particles 
outside the vacuum pipe, respectively.
The most downstream GTK station at $\mathrm{Z}= 102.4$~m marks the beginning of a 117~m long vacuum tank evacuated to a pressure of $10^{-6}$~mbar.

Momenta and directions of charged particles 
are measured by a magnetic spectrometer (STRAW).
The STRAW, comprising two pairs of straw chamber stations on either side of a
dipole magnet, measures momentum-vectors. The resolution of the momentum $p$ expressed in GeV$/c$ is 
$\sigma_p / p = (0.30 \oplus 0.005 \times p)\%$. 
The ring-imaging Cherenkov counter (RICH) is not used in the present analysis.
Two scintillator hodoscopes (CHOD and NA48-CHOD), consisting of a matrix of tiles and two orthogonal planes
of slabs, provide time measurements with 600 and 200~ps resolution, respectively.
Particle identification is provided by a quasi-homogeneous liquid
krypton electromagnetic calorimeter (LKr), two hadronic calorimeters (MUV1,2), and a muon detector (MUV3) just downstream of a 80~cm thick iron absorber. A
photon veto system includes the LKr, twelve ring-shaped
lead-glass detectors (LAV) and small angle calorimeters (IRC and SAC).
Synchronous energy deposits in nearby LKr cells are grouped into clusters.
The LKr resolution of the energy $E$ expressed in GeV   
is $\sigma_E / E = (4.8/\sqrt{E}\oplus11/E\oplus0.9)\%$. The LKr spatial and time
resolutions are 1~mm and between 0.5 and 1~ns, respectively, depending on the
amount and type of energy released.

\subsection{Data sample}
\label{sec:data}
Three trigger lines are employed during beam-dump operation. Two of them are used to identify charged particles: Q1, triggered by events with at least one signal in the CHOD and downscaled by a factor of 20; H2, triggered by events with at least two in-time signals in different tiles of the CHOD.
The third trigger line, the Control trigger, is used to identify  both charged and neutral particles.
The Control trigger requires a total energy above 1~GeV in the LKr, with one or more reconstructed clusters. More details on the trigger can be found in~\cite{na62trigperf,tdaq}.

The attenuation by the TAX allows the proton beam to be operated at a rate of 
$6.6\times10^{12}$ protons per spill of 4.8 seconds effective duration, equivalent to 1.7 times the intensity of the standard operation.
At this intensity, the 
rates of Control, downscaled Q1, and H2 triggers are 4, 14, and 18~kHz, respectively.

\section{Signal simulation}
\label{sec:signalMC}
Monte Carlo (MC) simulations of particle interactions with the detector and its response are performed using a software package based on the \verb|GEANT4| toolkit~\cite{GEANT4}. The response of the trigger lines is also emulated. 

After a proton interaction in the TAX, $A^\prime$ emission can proceed via a bremsstrahlung process or in a decay of secondary mesons. 
Bremsstrahlung production is understood in the generalized Fermi-Weizs{\"a}cker-Williams approximation from the scattering process~\cite{BlumleinBrunner13}
\begin{equation}
\label{eq:brems}
    \gamma^\ast p \to A^\prime p^\prime,
\end{equation}
where the virtual photon $\gamma^\ast$ is exchanged between the incoming proton $p$ and a nucleus ($N$), leading to a scattered proton $p^\prime$ and a dark photon $A^\prime$ in the final state. 
The production chain via meson decays can be summarized as 
\begin{equation}
\label{eq:meson}
    p N \to M X\mbox{, where }M = \pi^0\mbox{, }\eta^{(\prime)}\mbox{, }\rho\mbox{, }\omega\mbox{, }\phi,
\end{equation}
followed by
\begin{equation}
\label{eq:mesonTypes}
\begin{array}{llll}
    M \to & \gamma A^\prime & \mbox{ for } M = & \pi^0\mbox{, }\eta^{(\prime)}\mbox{;} \\
    M \to & \pi^0 A^\prime & \mbox{ for } M = & \eta^\prime\mbox{, }\rho\mbox{, }\omega\mbox{, }\phi\mbox{;} \\
    M \to & \eta A^\prime & \mbox{ for } M = & \rho\mbox{, }\omega\mbox{, }\phi.
\end{array}
\end{equation}
The \verb|PYTHIA| 8.2 generator~\cite{pythia} is used to model meson production. 
The differential cross-sections predicted by the simulation have been 
validated against available data~\cite{Dobrich:2019dxc}  
and agree to within 20\% or better in the full kinematic range.

Simulations of $A^\prime$ 
production and decay are used to evaluate the acceptance, the selection efficiency and other properties of the expected signal. 
 For each production mechanism, bremsstrahlung or meson decay, two decay modes, $A^\prime\to e^+e^-$ and $A^\prime\to\mu^+\mu^-$, are considered, with $A^\prime$ masses in the range 5--700~MeV$/c^2$ in 5-MeV$/c^2$ steps.
 The $A^\prime$ is constrained to decay in the volume $102<\mathrm{Z}<180$~m, with a decay path sampled from a flat distribution. At least $1.2\times10^5$ events are simulated for each mass value, production mechanism, and decay mode.

The expected dark photon yield for each value of the mass and coupling constant is expressed as:
\begin{equation}
\label{eq:DPYield}
N_\mathrm{exp} = N_p \times \mathrm{P}(pN \rightarrow A^\prime) \times \mathrm{P}_\mathrm{D} \times \mathrm{BR}(A^\prime \rightarrow \ell^+\ell^-)  \times A_\mathrm{sel},
\end{equation}
where
\begin{itemize}
    \item $N_p=1.4 \times 10^{17}$ is the number of protons dumped on TAX;
    \item $\mathrm{P}(pN \rightarrow A^\prime)$ is the $A^\prime$ production probability per proton: depending on the production mechanism, it accounts for the bremsstrahlung cross-section or the multiplicity of each meson type times the expected decay branching ratio quoted in eq.~(\ref{eq:mesonTypes});
    \item $\mathrm{P}_\mathrm{D}$ is the probability for the dark photon to decay within the range $102<\mathrm{Z}<180$~m, which depends on the $A^\prime$ lifetime and three-momentum distribution; 
    \item $\mathrm{BR}(A^\prime \rightarrow \ell^+\ell^-)$ is the branching ratio of the $A^\prime$ decay into a lepton pair;
    \item $A_\mathrm{sel}$ is the combined selection and trigger efficiency defined as:
    \begin{equation}
      \label{eq:DPefficiency}
      A_\mathrm{sel} = \left.{\textstyle\sum\limits_\mathrm{selected}}w_j \right / 
      {\textstyle\sum\limits_\mathrm{simulated}}w_i,
    \end{equation}
    where the sums run over the selected events and all simulated events. The weights $w_i$ are used to correct for the flat distribution of the $A^\prime$ decay paths $D_i$ sampled at generation level, and depend on the $A^\prime$ mean decay length $\lambda_i$:
    \begin{equation}
      \label{eq:DPLifetimeWeight}
      w_i = \frac{1}{\lambda_i}\,\, e^{- \frac{D_i}{\lambda_i}}.
    \end{equation}
\end{itemize}
A geometrical selection, which requires that the $A^\prime$ decays in the range $105<\mathrm{Z}<180$~m and its daughters are within the LKr active region, is used to compute $A_\mathrm{sel}$ in eq.~(\ref{eq:DPYield}). 

\begin{figure}[!ht]
\centering
\begin{tabular}{cc}
\includegraphics[width=.48\linewidth]{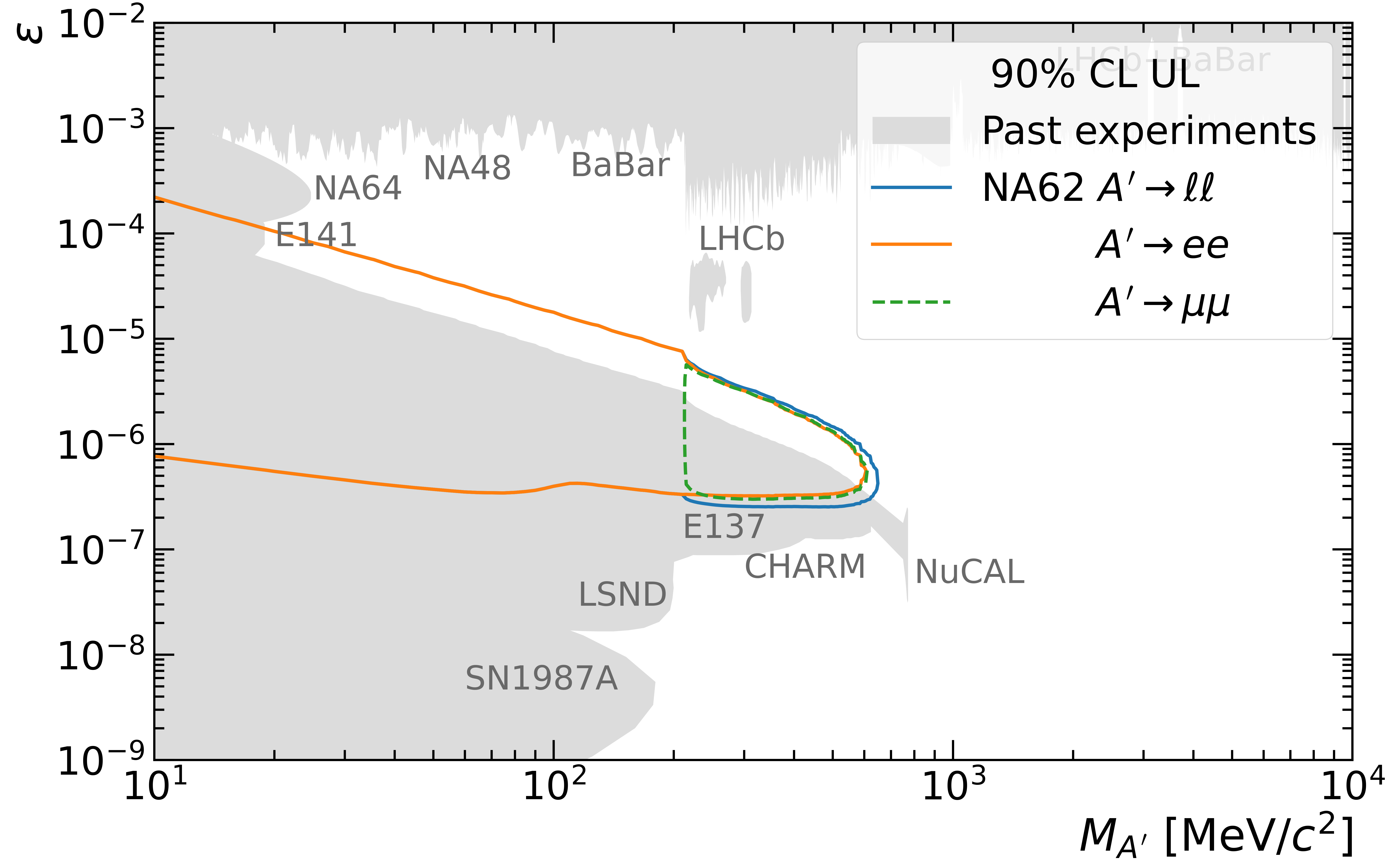}&
\includegraphics[width=.48\linewidth]{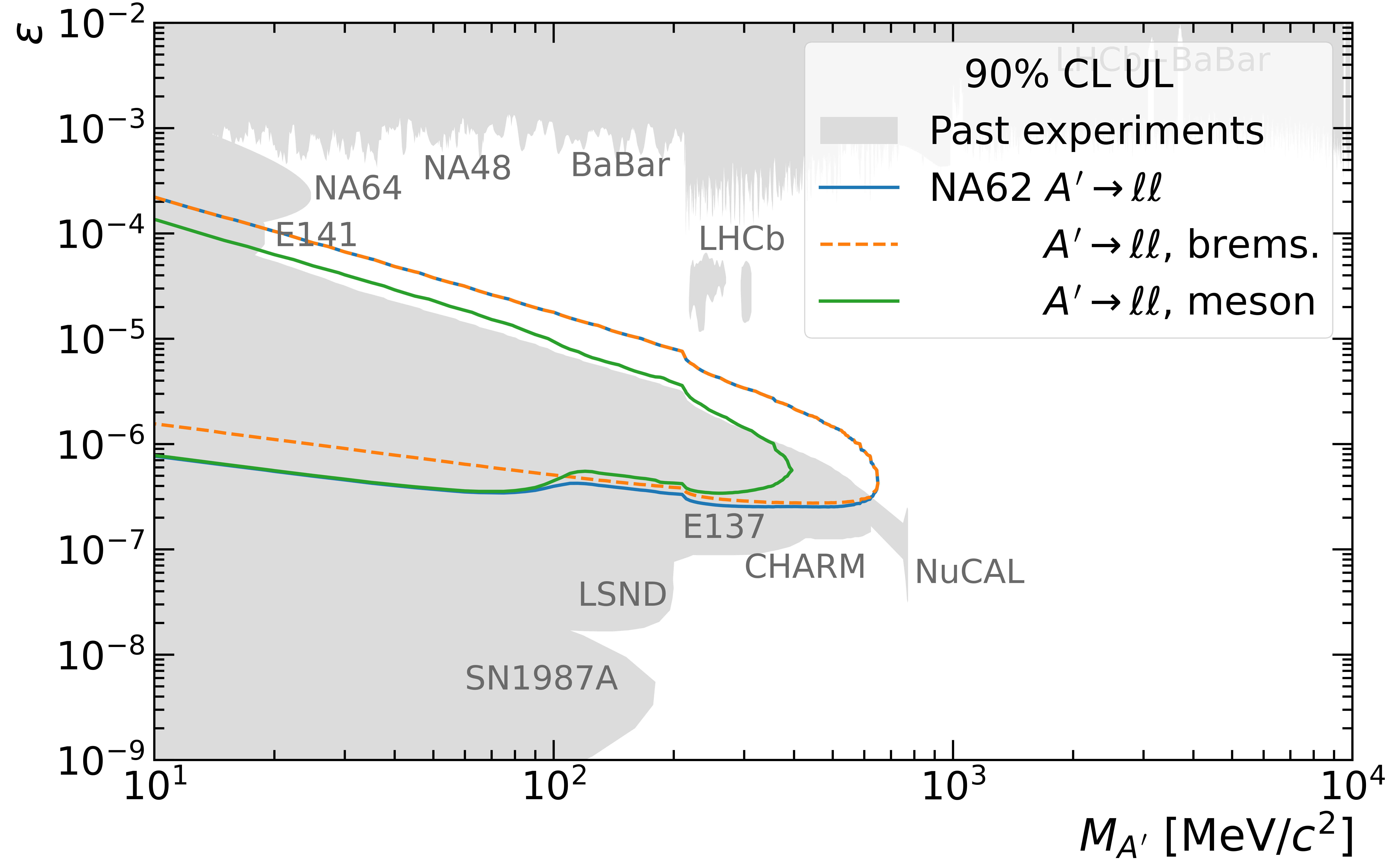}\\
\end{tabular}
\caption{\label{fig:DPSensitivityOnlyGeom} Regions excluded at 90\% CL assuming zero events observed in the absence of background for meson decays or bremsstrahlung $A^\prime$ production, separated by decay mode (left panel) and by production mode (right panel). 
The grey underlying excluded regions are obtained using the DarkCast package~\cite{Darkcast} and results from ref.~\cite{PBCSMWG}.}
\end{figure}

The resulting 90\% confidence level (CL) excluded region assuming zero events observed in the absence of background is shown in figure~\ref{fig:DPSensitivityOnlyGeom}. 
For masses above 215~MeV$/c^2$, the expected exclusion region from $\mu^+\mu^-$ decays is only marginally smaller than including both di-lepton modes. The yield of bremsstrahlung events exceeds that from meson decays due to the production cross-section
and the hardness of the spectra.  Therefore, the uncertainty on the meson production cross section is a sub-leading contribution for the present analysis.

\section{Event selection}
\label{sec:analysis}
Events triggered by the H2 condition are used for the signal search. 
A good quality track, reconstructed by the STRAW spectrometer, must satisfy the following requirements: momentum in excess of 10~GeV$/c$; downstream extrapolation within the geometrical acceptance of the NA48-CHOD, CHOD, LKr, MUV1, MUV2, MUV3 detectors, and within the inner aperture of the last LAV station; extrapolated positions at the front face of the first STRAW chamber and the LKr isolated from the other tracks; upstream extrapolation within the geometrical acceptance of the ANTI0; spatial association to an in-time CHOD signal. The track time is defined as the time of the associated NA48-CHOD signal  if present, otherwise of the associated CHOD signal, and must be within 5~ns of the trigger time.

An LKr cluster located within 50~mm of the track impact point and within 6~ns of the track time is associated to the STRAW track.  A MUV3 signal found within a momentum-dependent search radius around the track impact point and within 5~ns of the track time is associated to the STRAW track. A signal from MUV3 must only be associated to one STRAW track. 

Particle identification (PID) relies on the ratio $E/p$ of the LKr cluster energy associated ($E$) to the STRAW track momentum ($p$):
\begin{itemize}
    \item $\mu$ PID: zero or one associated LKr cluster with $E/p < 0.2$ and exactly one associated MUV3 signal;
    \item $e$ PID: one associated LKr cluster with $0.95<E/p<1.05$ and no associated MUV3 signal;
    \item $\pi$ PID: one associated LKr cluster with $0.2<E/p<0.9$ and no associated MUV3 signal.
\end{itemize}
Exactly one two-track vertex should be present in the event, reconstructed by extrapolating STRAW tracks backwards accounting for the residual magnetic field in the vacuum tank. The vertex Z coordinate must lie in the range 105--180~m. No requirement on the total charge at the vertex is imposed. The mean time of the two tracks defines the reference time. 
Vertices composed of oppositely charged tracks and $\mu$--$\mu$ PID assignments are considered as $A^\prime\to\mu^+\mu^-$ candidates.
No signal from any LAV station must be present within 10~ns of the reference time to reduce the background due to secondary interactions in the material.

\begin{figure}[t]
\begin{center}
\includegraphics[width=.6\textwidth]{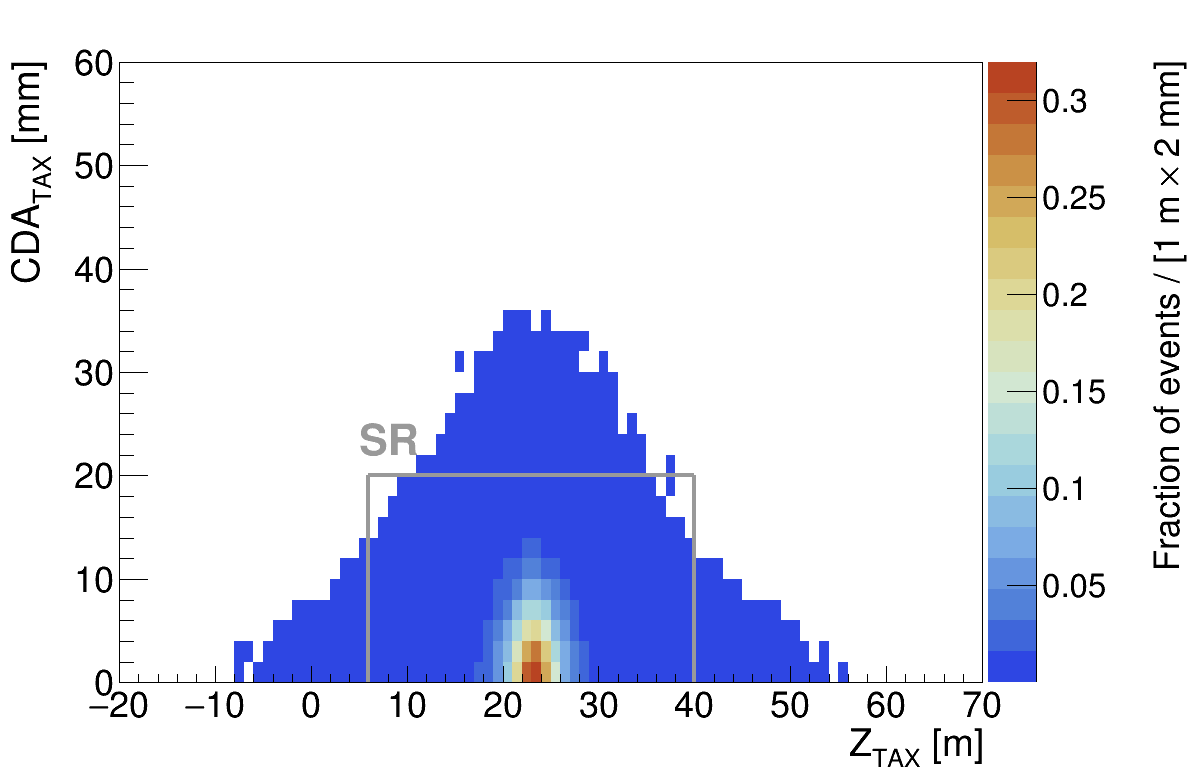}
\caption{\label{fig:CDA_Vtx_TAXMC} Distance of closest approach $\mathrm{CDA}_\mathrm{TAX}$ vs longitudinal coordinate of the point of minimum approach $\mathrm{Z}_\mathrm{TAX}$ for simulated signal events. 
The signal region defined by eq.~(\ref{eq:SR}) is shown inside the rectangular contour.}
\end{center}
\end{figure}

 The position of the $A^\prime$ production point is evaluated as the point of closest approach, $P_\mathrm{CDA}$, between the dark photon line of flight (defined by the vertex position and the sum of the three-momenta at the vertex) and the proton beam line (defined by the average impact point on the dump, eq.~(\ref{eq:beamImpact}), and parallel to the Z axis). 
 The distance of closest approach $\mathrm{CDA}_\mathrm{TAX}$ is shown in figure~\ref{fig:CDA_Vtx_TAXMC} as a function of the longitudinal coordinate $\mathrm{Z}_\mathrm{TAX}$ of $P_\mathrm{CDA}$ for simulated signal events. The $\mathrm{Z}_\mathrm{TAX}$ distribution has a mean value of 23~m with a rms width of 5.5~m. The rms width of the $\mathrm{CDA}_\mathrm{TAX}$ distribution is 7~mm. The signal region (SR) is defined as 
\begin{equation}
\label{eq:SR}
\mathrm{SR:}\,\,6<\mathrm{Z}_\mathrm{TAX}<40~\mathrm{m}\,\,\mathrm{ and }\,\,\mathrm{CDA}_\mathrm{TAX}<20~\mathrm{mm,}
\end{equation}
and the validation region (VR) is defined as
\begin{equation}
\label{eq:CR}
\mathrm{VR:}\,\,-4<\mathrm{Z}_\mathrm{TAX} <50~\mathrm{m}\,\,\mathrm{ and }\,\,\mathrm{CDA}_\mathrm{TAX}<150~\mathrm{mm},
\mbox{ excluding SR.}
\end{equation}
The VR is used for validation of the background estimates with the data, allowing the unmasking of the SR if a satisfactory agreement is found.

\section{Background determination}
The evaluation of the expected background would require the simulation of about $N_p=10^{17}$, which is technically too demanding.
A combination of data-driven and MC methods was developed to overcome this difficulty.

\begin{figure}[t]
\centering
\includegraphics[width=.7\linewidth]{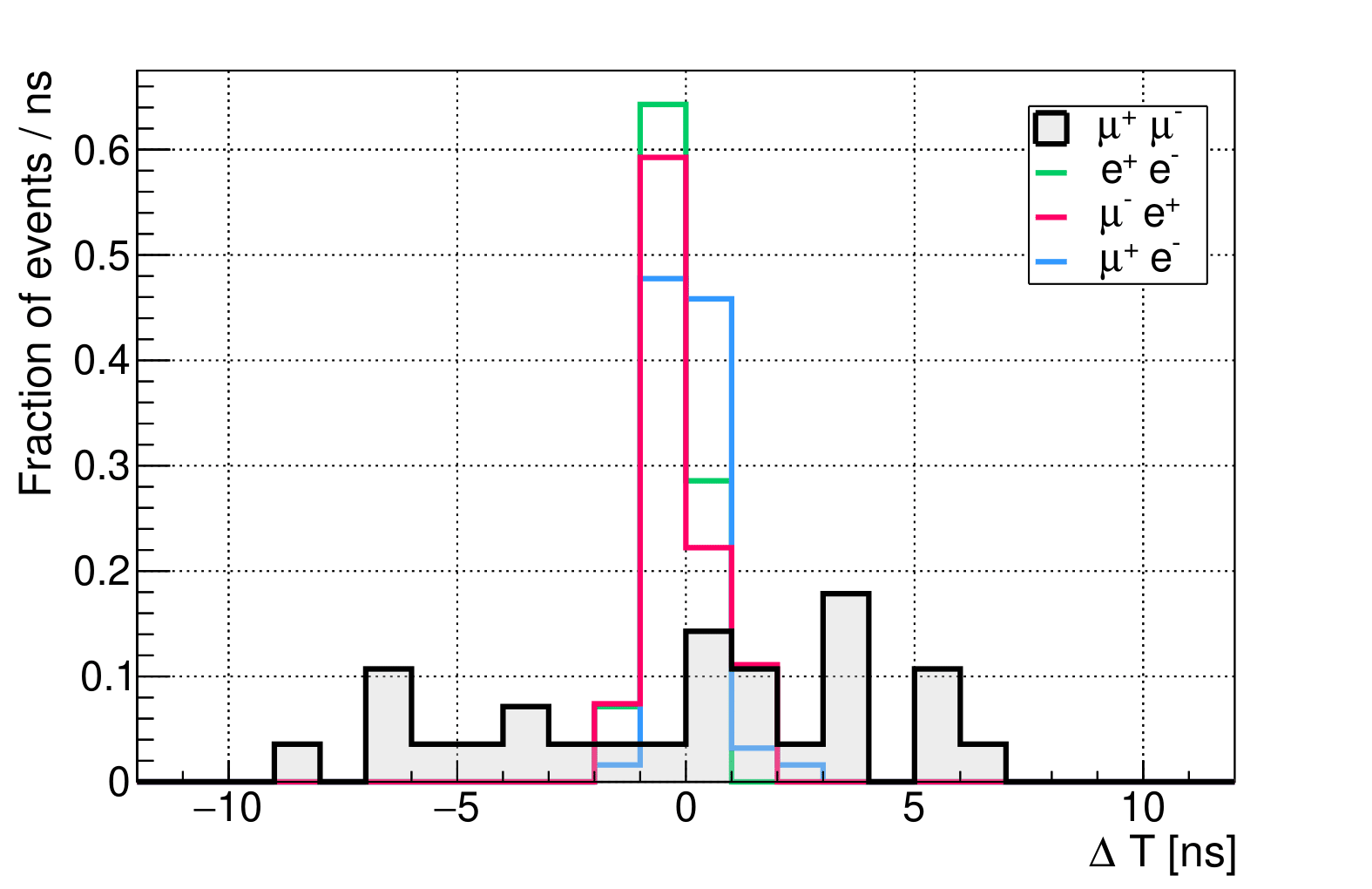}
\caption{\label{fig:DeltaTime} Time difference between the two selected tracks for various PID combinations: $\mu^+\mu^-$ (black), $e^+e^-$ (green), $\mu^-e^+$ (red), $\mu^+e^-$ (light blue).}
\end{figure}

The distribution of the time difference between the two selected tracks, inverting some of the selection criteria, is exploited to give indications about the origin of the expected background. The following data side bands are considered:
\begin{itemize}
    \item Opposite-charge vertices with $e$--$e$ or $\mu$--$\mu$ PID, outside  
     the signal and validation regions; 
    \item Same-charge vertices with $\mu$--$\mu$ PID, both outside and within the signal or validation regions;
    \item Same- or opposite-charge vertices with $e$--$\mu$ PID, both outside and within the signal or validation regions.
\end{itemize}
The time difference distributions, shown in figure~\ref{fig:DeltaTime}, indicate that vertices with at least one electron or positron are formed mostly by in-time tracks: this ``prompt" background can be explained by secondary interactions of incoming muons within the material traversed. In contrast, di-muon vertices formed by unrelated tracks randomly paired produce a ``combinatorial” background with a uniformly distributed time difference.

\subsection{Prompt background}
\label{SSec:InTimeBKGMC}
In the available data set, $5 \times 10^{9}$
halo muons are in the acceptance of the CHOD, LKr, and MUV3 detectors. 
The prompt background originates from interactions of halo muons in the material traversed upstream of or within the decay volume.  
The main prompt background mechanism is muon-nucleus inelastic production of a hadron, usually a charged pion, followed by an in-flight decay to a muon. Two in-time muons are then present in the event. 

Two approaches to the simulation of the muon flux after proton interactions in the TAX have been exploited. The first method
consists of the simulation of a limited number of proton interactions and the parameterisation of the muon kinematics at the TAX exit plane. This parameterisation is then used for simulation~\cite{Rosental}. The second method
enhances the proton-induced muon production to increase the simulation efficiency~\cite{JHEPMC}. However, neither approach led to satisfactory results due to: the limited knowledge of the pion/kaon cross-sections for forward production and for quasi-elastic scattering in 
TAX nuclei; the uncertainties in the multiple scattering treatment, particularly within the iron yokes of the  
beam line magnets. Moreover, both methods require an oversampling of the resulting  halo muons of the order of a thousand times to achieve a number of events equivalent to $N_p=10^{17}$, potentially inducing non-physical correlations. 

To overcome these issues, a backward MC simulation (BMC) fed with real data is
used.
The input consists of a set of distributions from single tracks with $\mu$ PID:
X,Y coordinates and three-momentum components measured at a reference plane ($\mathrm{Z}=180$~m) upstream of the STRAW spectrometer. 
\verb|PUMAS|~\cite{PUMAS}, a standalone tool used in muography studies and interfaced with \verb|GEANT4|, propagates each muon backward, increasing its energy according to the amount of material traversed, until reaching the upstream face of the B5 magnet at $\mathrm{Z} = 92$~m.
The result is a sample of muons, which is expected to reproduce the experimental
distributions.
To validate the method, the sample of muons from BMC is
 input into the NA62 standard MC simulation based on \verb|GEANT4| and the results at the reference plane are compared to the original data.
Disagreements can be explained by the different treatments of multiple scattering  in \verb|PUMAS| and \verb|GEANT4| and the asymmetric distribution of the energy loss, which induces tails at high momenta.
To correct for such biases, a weight depending on the track momentum and its radial position at the B5 magnet plane is assigned to each muon track.
A systematic uncertainty of $50\%$ in results obtained using these simulations is derived from the comparison between data and MC distributions of angles and positions in the transverse plane.

Technical limitations for the full halo muon simulation remain, particularly because of muon-induced showers downstream of the LKr, $\mathrm{Z}_\mathrm{LKr} = 241$~m.
Therefore, the MC simulation is split into two stages. 
All particles are propagated from the B5 magnet to the STRAW spectrometer downstream plane ($\mathrm{Z}_\mathrm{STRAW}=219$~m). 
Events are then kept for further propagation if either ({\it a}) one $e^{\pm}/\gamma/\pi^{\pm}/p/n/K^{\pm}/K^0_L$   
 with a momentum above $1$~GeV$/c$ or ({\it b}) at least two muons, regardless of their charge, reach $\mathrm{Z}_\mathrm{STRAW}$.

A number of events equivalent to $N_p=0.67\times10^{17}$ ($8.37\times 10^{15}$) is generated using the condition {\it a} ({\it b}). 
Pions produced by muon interaction can decay 
at $\mathrm{Z}<\mathrm{Z}_\mathrm{STRAW}$ (``$\mu$-$\mu$" background) or at $\mathrm{Z}_\mathrm{STRAW}<\mathrm{Z}<\mathrm{Z}_\mathrm{LKr}$ (``$\mu$-$\pi$'' background). 
To increase the statistics of the $\mu$-$\pi$
component, events are oversampled forcing the pion decay to a muon before reaching the LKr.
An additional background component (``Other'') is  due to: $K^\pm$ production 
followed by a decay to muons; muon hard ionisation with emission of $e^\pm$ interacting before reaching the LKr. In the latter case, the emitted particles can be accidentally associated to a MUV3 in-time signal from the original muon.

\begin{figure}[t]
\begin{center}
\includegraphics[width=.7\linewidth]{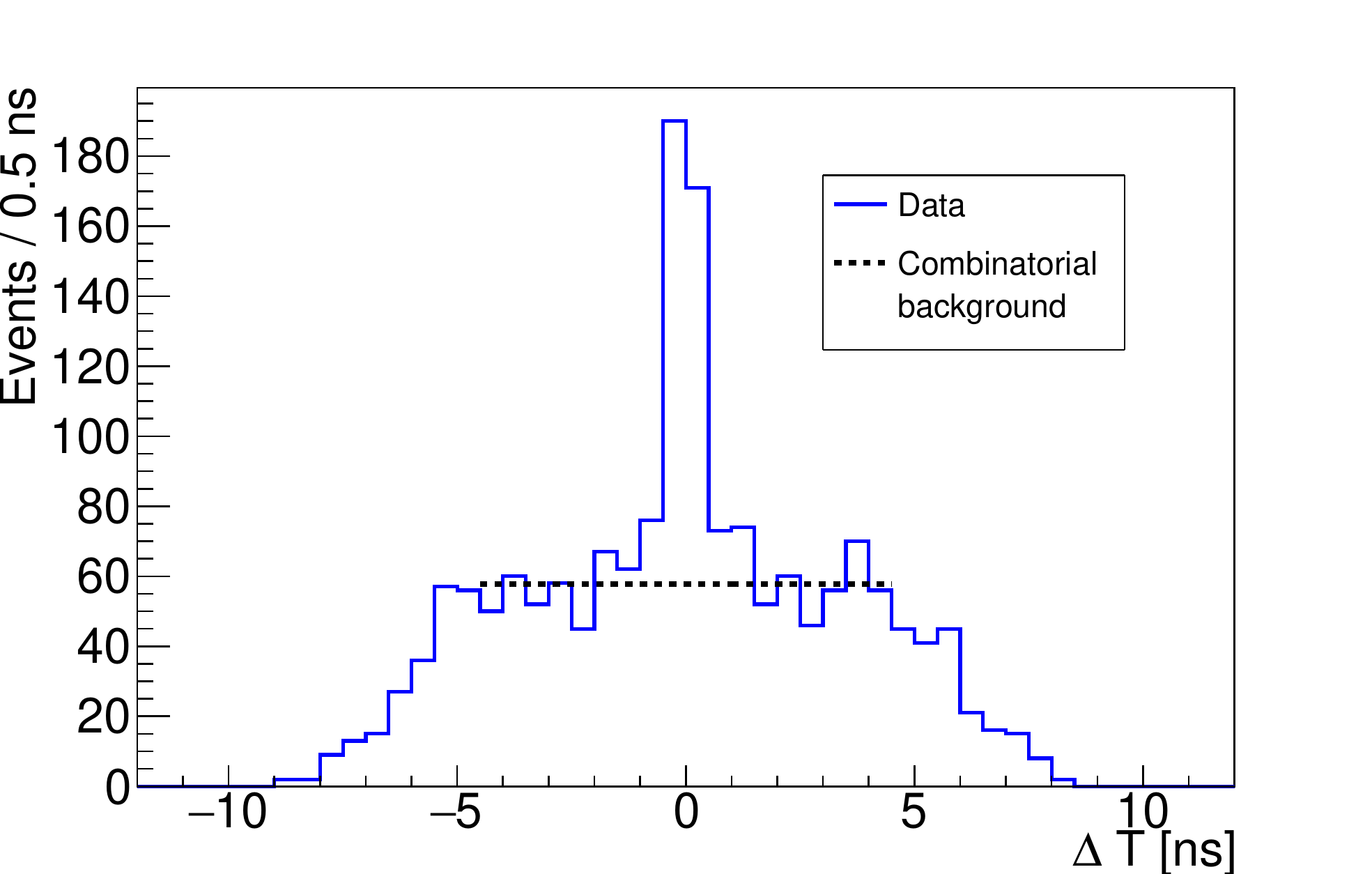}
\caption{\label{fig:mumuBeforeLAVDeltaTime} Time difference between the two tracks selected as $\mu^+\mu^-$, without the LAV veto condition (section~\ref{sec:analysis}) and excluding vertices in the VR or SR.}
\end{center}
\end{figure}

The expected and observed numbers of events satisfying the selection without the LAV veto condition are compared, excluding the signal and validation regions. The distribution of the two-track  time difference for $\mu^+\mu^-$ data events is shown in figure~\ref{fig:mumuBeforeLAVDeltaTime}.  
The expected combinatorial background is evaluated from the side bands ($1 < | \Delta T | < 4.5$ ns) assuming a uniform
distribution and it is subtracted to obtain the prompt component ($| \Delta T | < 1$ ns).
The prompt component amounts to $270\pm27_\mathrm{stat}$ events. From the simulation, $141\pm66_\mathrm{stat}\pm71_\mathrm{syst}$ prompt-background events are expected.
The data/MC ratio, $1.91\pm0.91_\mathrm{stat}\pm0.95_\mathrm{syst}$, is used to scale 
the predictions from the MC simulation.

\begin{table}
	\centering
	\caption{Expected numbers of prompt-background events for $N_p=1.4\times 10^{17}$ obtained from simulations.
    The signal selection is applied, and events in the SR or {VR} are excluded. 
    The uncertainties quoted are statistical; the second uncertainty in the last column is systematic.}
    \vspace{0.2cm}
	\begin{tabular}{|c|c|c|c|}
		\hline
		$\mu$-$\mu$ & $\mu$-$\pi$ & Other & Total\\
		\hline
        0.235 $\pm$ 0.177 & 0.038 $\pm$ 0.019 & 0.004 $\pm$ 0.003 & 0.28 $\pm$ 0.19 $\pm$ 0.20 \\ \hline
	\end{tabular}
	\label{tab:prompt_newSel}
\end{table}

\begin{figure}[t]
\begin{center}
\includegraphics[width=.6\linewidth]{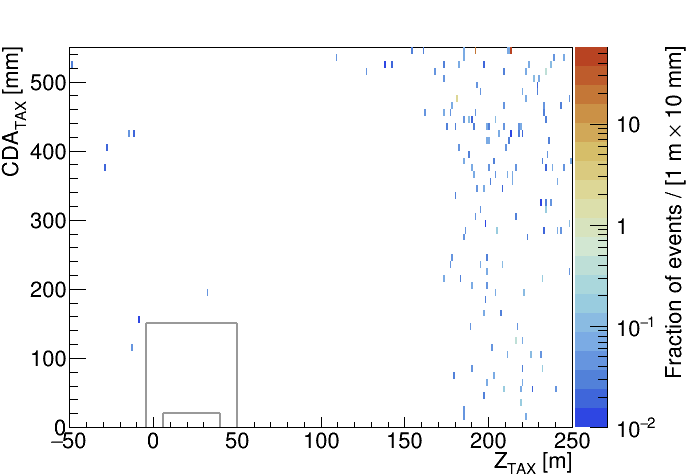}
\caption{\label{fig:CDA_Vtx_TAXMCPrompt} $\mu^+\mu^-$ expected background distribution of the prompt component before the LAV veto condition, in the $(\mathrm{Z}_\mathrm{TAX}\mbox{, }\mathrm{CDA}_\mathrm{TAX})$ plane. The rectangles are the external contours of SR and VR regions.}
\end{center}
\end{figure}

The expected numbers of events due to the prompt background with the LAV veto condition applied are given in table~\ref{tab:prompt_newSel}. The  distribution of the prompt background before the LAV veto condition in the  ($\mathrm{Z}_\mathrm{TAX}\mbox{, }\mathrm{CDA}_\mathrm{TAX}$) plane is exploited to evaluate the fraction of background events in the VR ($\eta_\mathrm{VR}$). As shown in figure~\ref{fig:CDA_Vtx_TAXMCPrompt}, no simulated  events are observed in the VR. At 90\% CL, $\eta_\mathrm{VR}<1.6\%$. The corresponding upper limit on the number of expected events in the VR is 0.004.

A possible prompt-background contribution produced by secondary interactions in the collimators or magnets preceding the decay volume is also investigated.  In the $(\mathrm{Z}_\mathrm{TAX}\mbox{, }\mathrm{CDA}_\mathrm{TAX})$ plane, the distribution of the upstream-produced prompt background does not significantly differ from that of figure~\ref{fig:CDA_Vtx_TAXMCPrompt}. A conservative estimate establishes an upper limit of 0.069 expected events in the VR at 90\% CL.

\subsection{Combinatorial background}
\label{sec:combinatorial}
A control data sample is used to evaluate the combinatorial background. 
The control sample consists of events satisfying the Q1 but not the H2 trigger conditions,
to avoid any overlap with the signal selection.
Events with a single STRAW track in time with the trigger are selected. The requirements of track quality, association with downstream detectors, and $\mu$ PID are applied as in the signal selection. 

The selected single tracks are paired, simulating a random coincidence within 10~ns in the same event. The vertex reconstruction is performed as in the signal selection. 
Each simulated track pair is weighted to account for the time window required by the signal selection, the spill duration, the downscale factor of the Q1 trigger and the efficiency for the H2 trigger given two tracks fulfilling the Q1 condition.
The relative systematic uncertainty in the event weight is 15\% and significantly outweighs the statistical uncertainty.

\begin{figure}[ht]
\centering
\includegraphics[width=.48\linewidth]{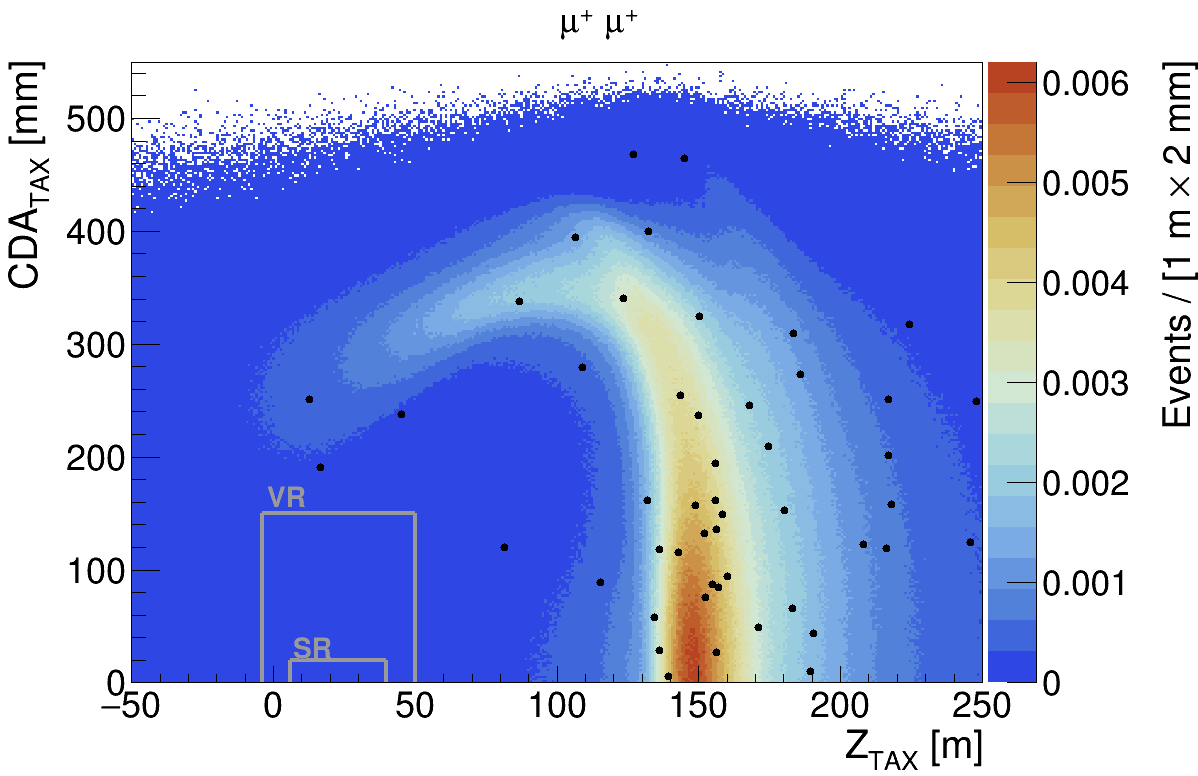}~
\includegraphics[width=.48\linewidth]{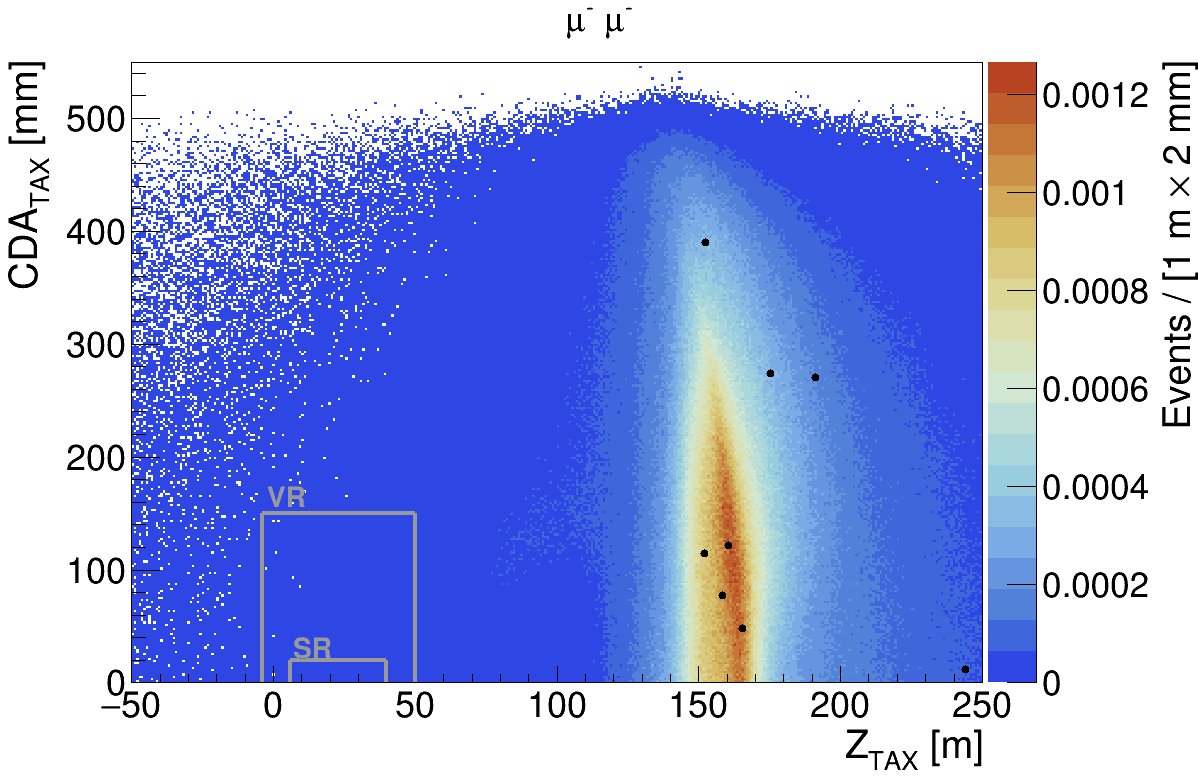}
\caption{\label{fig:MuMuBkg} Distributions of $\mathrm{CDA}_\mathrm{TAX}$ vs $\mathrm{Z}_\mathrm{TAX}$ for $\mu^+\mu^+$ (left) and $\mu^-\mu^-$ (right) events: expected combinatorial background  (colour-scale plot) and data events (black dots). Data events in the SR and VR  are not masked.}
\end{figure}

\begin{figure}[t]
\centering
\includegraphics[width=.6\linewidth]{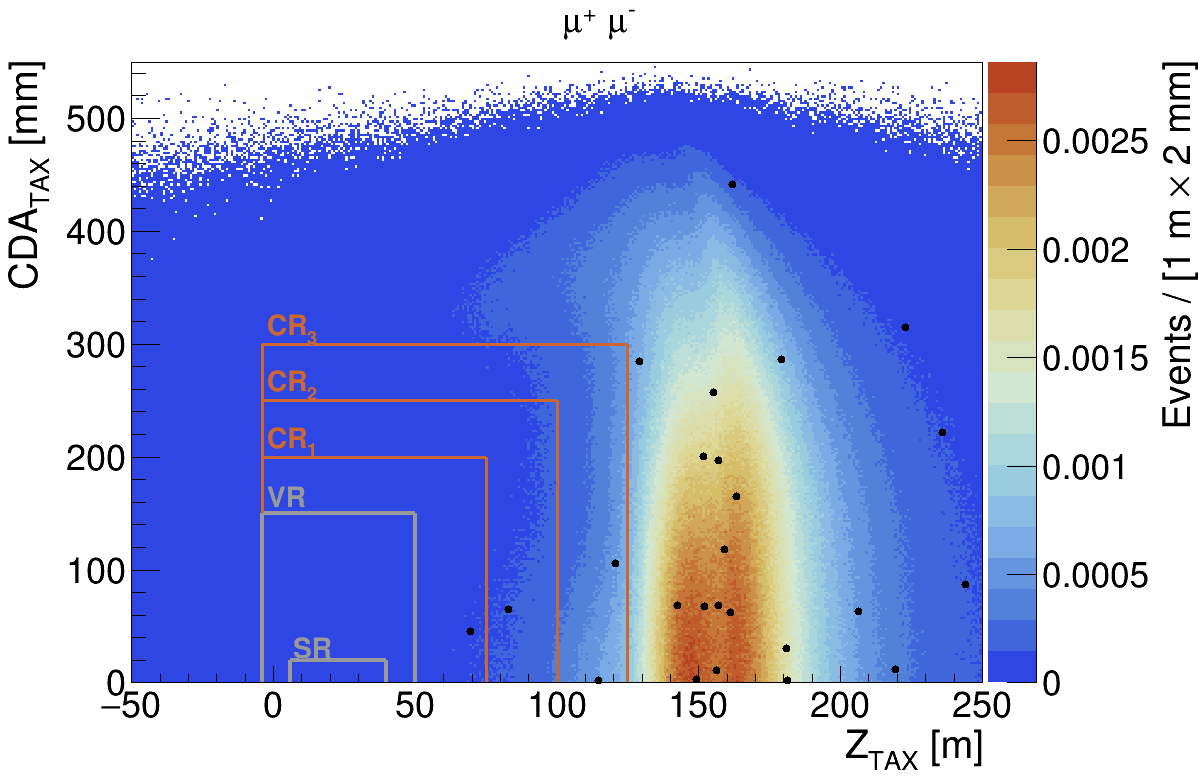}
\caption{\label{fig:MuMuOppositeSignBkg} Distribution of $\mathrm{CDA}_\mathrm{TAX}$ vs $\mathrm{Z}_\mathrm{TAX}$ for $\mu^+\mu^-$ events: expected combinatorial background  (colour-scale plot) and data events (black dots). Validation and signal regions are masked for data. Additional regions, CR$_{3,2,1}$, are not masked.}
\end{figure}

After weighting events, the distributions of $\mathrm{CDA}_\mathrm{TAX}$ vs $\mathrm{Z}_\mathrm{TAX}$ for $\mu^+\mu^+$ and $\mu^-\mu^-$ events are shown in figure~\ref{fig:MuMuBkg}. Data events are superimposed as full dots. 
In figure~\ref{fig:MuMuOppositeSignBkg}, $\mu^+\mu^-$ events are shown. 
Three control regions closer and closer to the VR and labelled as CR$_{3,2,1}$ are considered. 
A 
comparison between observed and expected numbers of events is shown 
in table~\ref{tab:DiMuonCombinatorialNewSel} and a good agreement is observed.
\begin{table}
    \caption{Numbers of expected di-muon events from combinatorial background ($N_\mathrm{exp}$), 
    numbers of observed data events ($N_\mathrm{obs}$), and 
    probabilities to obtain a likelihood L for data-MC compatibility equal or smaller than that corresponding to $N_\mathrm{obs}$ ($P_{\mathrm{L}\leq \mathrm{L}_\mathrm{obs}}$). The dominant uncertainty in $N_\mathrm{exp}$ is systematic.}
   \vspace{0.2cm}
    \centering
    \begin{tabular}{|c|c|c|c|c|}
        \hline
          PID & Region & $N_\mathrm{exp}$ & $N_\mathrm{obs}$ & 
          $P_{\mathrm{L}\leq \mathrm{L}_\mathrm{obs}}$ \\
        \hline

                         & Outside { VR} & $62.5\pm9.4$ & 53 & 0.46\\
        $\mu^+\mu^+$     & { VR} & $0.46\pm0.07$ &  0 &  1.0 \\
                         & SR & $0.040\pm0.006$ & 0  & 1.0 \\ \hline
                         & Outside { VR} & $9.1\pm1.4$ & 8  & 0.88\\
        $\mu^-\mu^-$     & { VR} & $0.050\pm0.007$ & 0  & 1.0\\
                         & SR & $0.0050\pm0.0007$ & 0  & 1.0 \\ \hline
                         & Outside {VR} & $30.9\pm4.6$ & 28  & 0.78 \\
                         & CR$_3$ & $2.00\pm0.30$ & 2  & 1.0\\
                         & CR$_2$ & $0.68\pm0.10$ & 1  & 0.48\\
        $\mu^+\mu^-$     & CR$_1$ & $0.34\pm0.05$ & 1  & 0.29\\ 
                         & CR$_{1+2+3}$ & $3.02\pm0.45$ & 4  & 0.56 \\ 
                         & { VR} & $0.20\pm0.04$ & -- & -- \\
                         & SR & $0.019\pm0.004$ & -- & -- \\ \hline

	   \hline 
    \end{tabular}
\label{tab:DiMuonCombinatorialNewSel}
\end{table}

For the $\mu^+\mu^-$ final state, an alternative evaluation of the combinatorial background  
is obtained by determining the data/MC scaling from same-sign events outside the VR: 61 events are observed in data, while $71.6\pm9.5$ are expected. The scale factor is lower than that used in the previous approach by 15\%, although consistent within the systematic error.  Using same-sign events for scaling allows a relative statistical uncertainty of 13\% and a negligible systematic contribution. The final estimate of the combinatorial background employs this alternative approach and is shown in table~\ref{tab:DiMuonCombinatorialAlternativeNewSel}.
\begin{table}
    \caption{Numbers of expected $\mu^+\mu^-$ events from combinatorial background ($N_\mathrm{exp}$), numbers of data events ($N_\mathrm{obs}$), and probabilities to obtain a likelihood L for data-MC compatibility equal or smaller than that corresponding to $N_\mathrm{obs}$ ($P_{\mathrm{L}\leq \mathrm{L}_\mathrm{obs}}$). The data/MC ratio for same-sign events is used to determine the MC scaling factor. The dominant uncertainty in $N_\mathrm{exp}$ is statistical.}
    \vspace{0.2cm}
    \centering
    \begin{tabular}{|c|c|c|c|}
        \hline
          Region  & $N_\mathrm{exp}$ & $N_\mathrm{obs}$ & 
          $P_{\mathrm{L}\leq \mathrm{L}_\mathrm{obs}}$ \\
        \hline
             Outside {VR} & $26.3\pm3.4$ & 28 &  0.74 \\
             {{CR$_3$}} & $1.70\pm0.22$ & 2 &  0.68\\
             {CR$_2$} & $0.58\pm0.07$ & 1 &  0.44\\
             {CR$_1$} & $0.29\pm0.04$ & 1 &  0.25\\ 
             {CR$_{1+2+3}$} & $2.57\pm0.33$ & 4 & 0.34 \\ \hline
             {VR} & $0.17\pm0.02$ & -- & -- \\
             SR & $0.016\pm0.002$ & -- & -- \\ \hline

	   \hline 
    \end{tabular}
\label{tab:DiMuonCombinatorialAlternativeNewSel}
\end{table}

\subsection{Background summary}
The estimates of the prompt and combinatorial backgrounds are displayed in table~\ref{tab:DiMuonNewSelTotalBkg}. 
The fraction of events within the SR is expected to be ten times smaller than within the VR, assuming a flat distribution in these regions.
The evaluations of the combinatorial background support this assumption, which is used for the prompt and upstream-prompt components.
The total expected number of background events is $0.016\pm0.002$, dominated by the combinatorial component.
Assuming a 90\% CL coverage and no signal, no observed events are expected in the data SR.
A five-sigma signal discovery for any mass $M_{A^\prime}$ would correspond to the observation of three or more signal candidates in the data SR in a window of $\pm3$ standard deviations of the mass.
\begin{table}[h]
    \caption{Summary of expected numbers of background events for the search of $A^\prime\to\mu^+\mu^-$ with the related uncertainty. The limits reported are defined with a 90\% CL.}
    \vspace{0.2cm}
    \centering
    \begin{tabular}{|c|c|c|c|}
        \hline
          Region & Combinatorial & Prompt & Upstream-prompt \\
        \hline
        { VR} & $0.17\pm0.02$   & $<0.004$ & $<0.069$ \\
        SR & $0.016\pm0.002$ & $<0.0004$ & $<0.007$ \\ \hline

	   \hline 
    \end{tabular}
\label{tab:DiMuonNewSelTotalBkg}
\end{table}

\section{Expected signal yield}
\label{sec:sensitivity}
The signal 
yield is obtained using eq.~(\ref{eq:DPYield}). 
The number of protons on TAX ($N_p$) is evaluated for each spill from the measurement of the beam flux which is provided by a titanium-foil secondary-emission monitor placed at the target location. 
The uncertainty in the $N_p$ measurement is derived from the operational experience of the secondary-emission monitors in various beam-line setups and is estimated conservatively to be 20\%. This figure is confirmed using data from the standard setup: the number of selected $K^+\to \pi^+\pi^+\pi^-$ decays agrees with the number expected from the measured proton flux to within $20\%$.

\begin{figure}[t]
\begin{center}
\includegraphics[width=.49\linewidth]{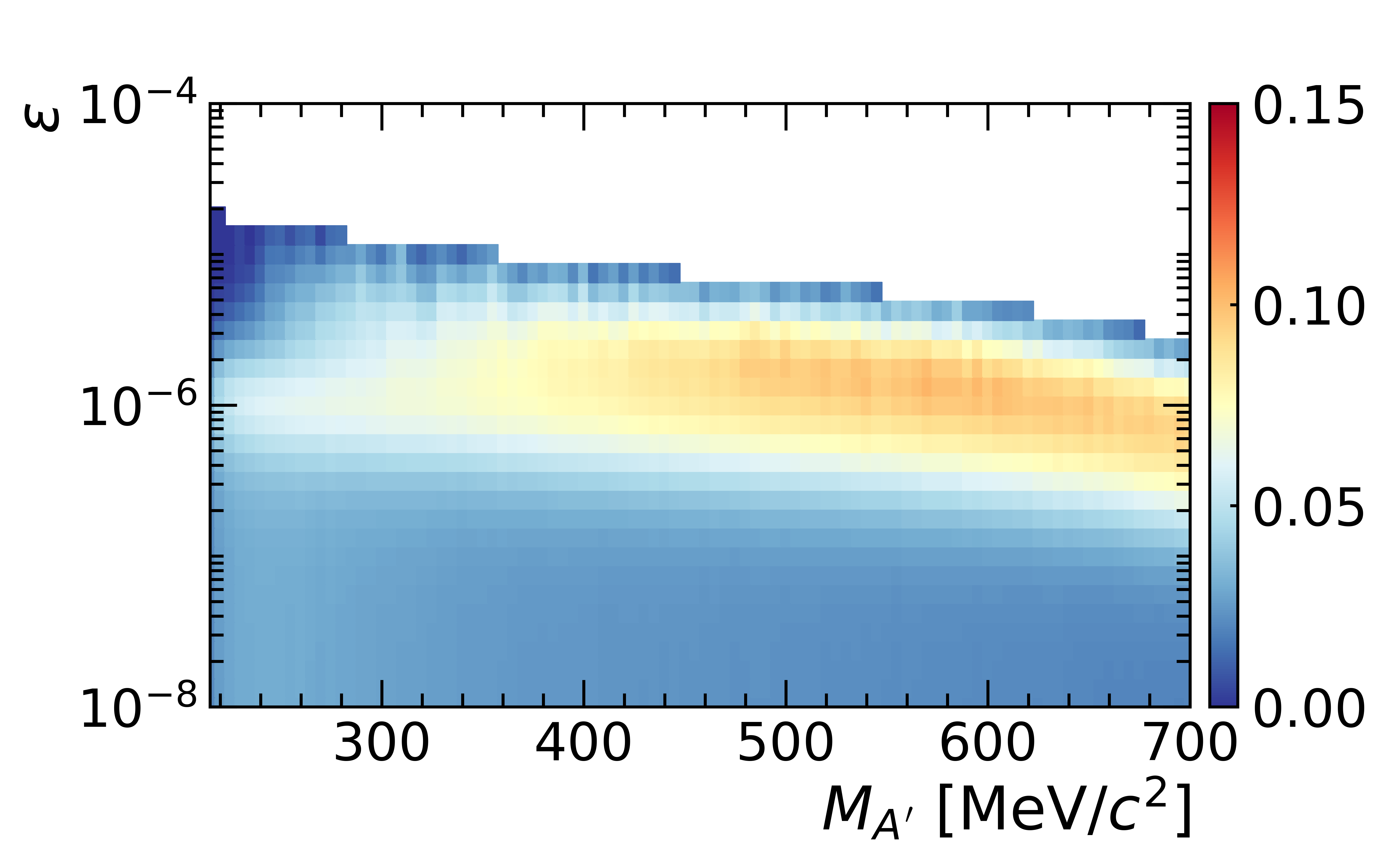}
\includegraphics[width=.49\linewidth]{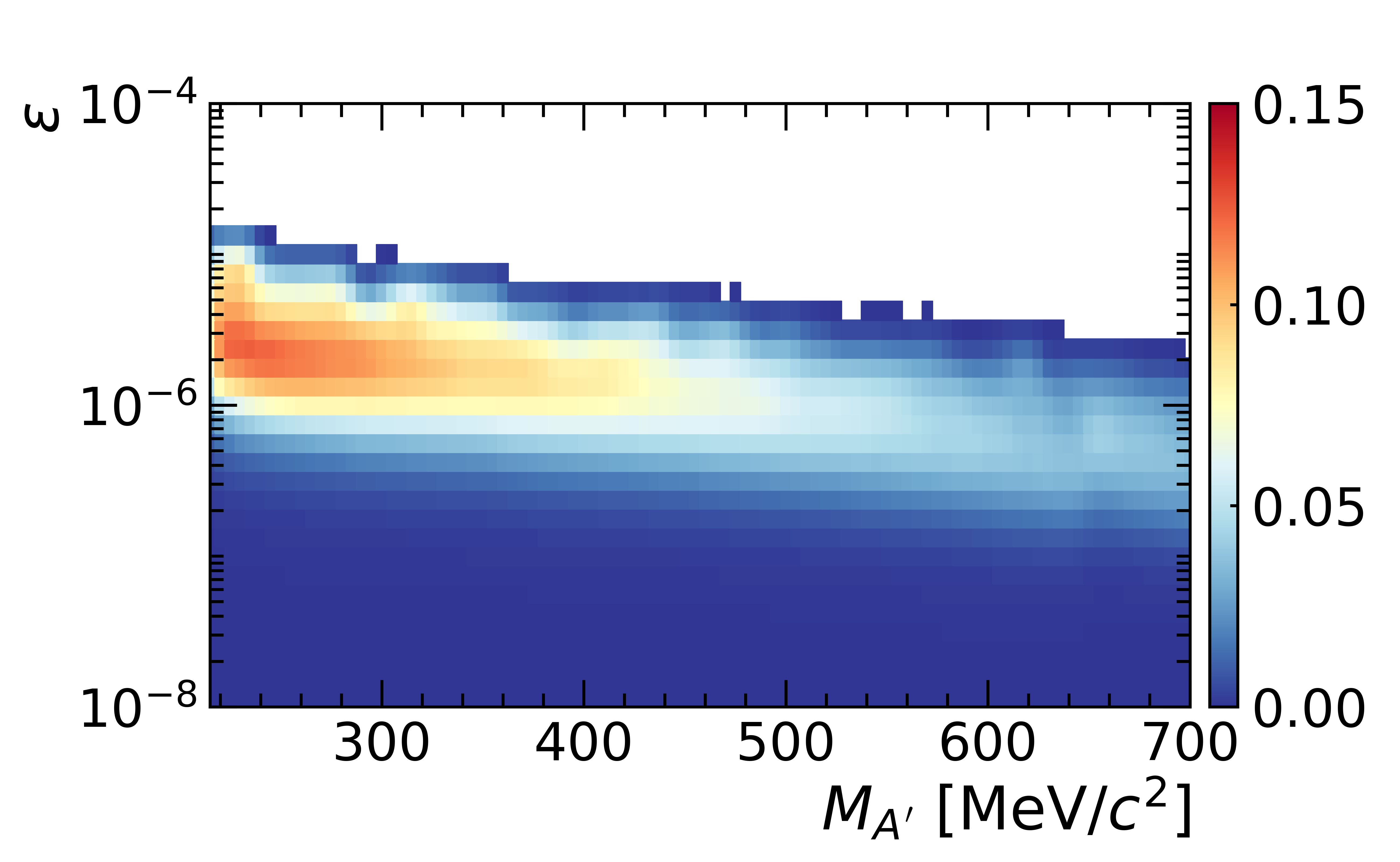}
\caption{\label{fig:effiplots} Selection and trigger efficiency for the $A^\prime\to\mu^+\mu^-$ signal, as a function of the $A^\prime$ mass and coupling constant. Left (right) panel refers to the bremsstrahlung (meson-decay) production mode.}
\end{center}
\end{figure}

The selection and trigger efficiencies are determined by simulation as a function of the assumed 
dark photon mass and coupling constant, separately for the bremsstrahlung and for the meson-decay production processes. The mass is varied from 215~MeV$/c^2$ to 700~MeV$/c^2$. The resulting efficiencies are shown in figure~\ref{fig:effiplots}. The trigger inefficiency is 2\%.
For any value of $M_{A^\prime}$, the maximum efficiency occurs at a given value of $\varepsilon$, because of two competing effects: for larger values of the coupling constant $\varepsilon$, the average $A^\prime$ momentum is higher to compensate for the lower $A^\prime$ lifetime at rest, leading to reduced di-muon opening angles and therefore lower reconstruction efficiency for tracks and vertices; for lower values of  $\varepsilon$, the $A^\prime$ lifetime at rest is longer and softer dark photons are selected, leading to a reduced acceptance of the $A^\prime$ decay products. 

A summary of the relative systematic uncertainties in the signal selection efficiency is given in table~\ref{tab:effiNewSel}. 
Each entry is determined independently using a combination of data control samples and simulation. 
The simulation entry is of statistical origin and represents a typical value, since it varies with the $A^\prime$ mass and coupling constant. 
The total relative uncertainty in the efficiency is below 3\%.
\begin{table}
    \caption{Relative uncertainties of the signal selection efficiency from the contributions considered.}
    \vspace{0.2cm}
    \centering
    \begin{tabular}{|l|c|}
        \hline
         Source & Uncertainty   \\
        \hline
        Track and vertex reconstruction  & $<0.1$\%   \\ 
        CHOD association       & 0.6\%  \\
        PID                    & 1.0\%  \\
        LAV veto condition  & 0.1\%  \\
        Extrapolation to the impact point  & 1.5\%   \\ 
        Trigger                & 0.5\%  \\
        Simulation            & 2.1\%  \\
	   \hline 
	   Total                   & 2.8\%  \\ \hline
    \end{tabular}
\label{tab:effiNewSel}
\end{table}

\begin{figure}[t]
\centering
\includegraphics[width=.49\linewidth]{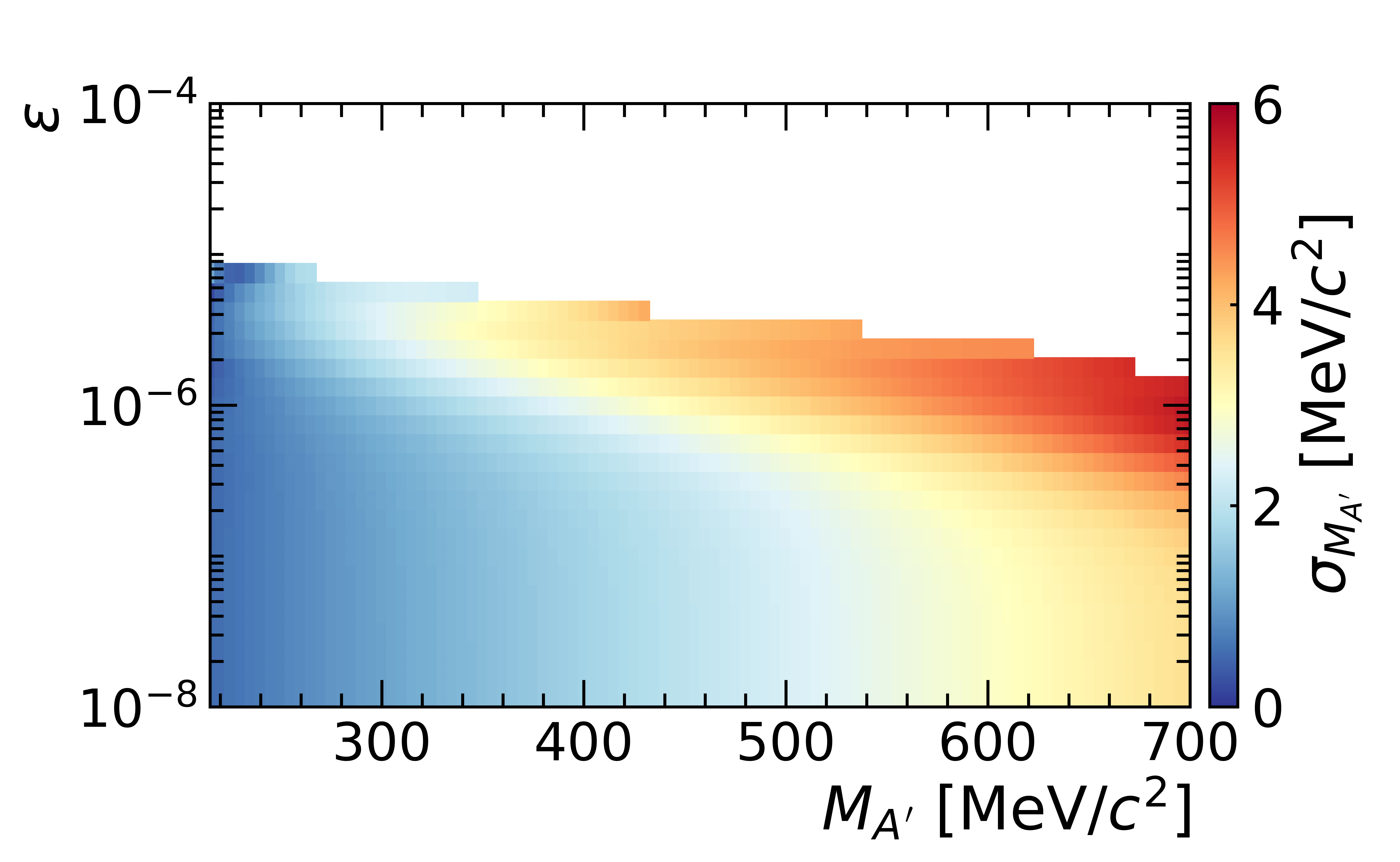}
\includegraphics[width=.49\linewidth]{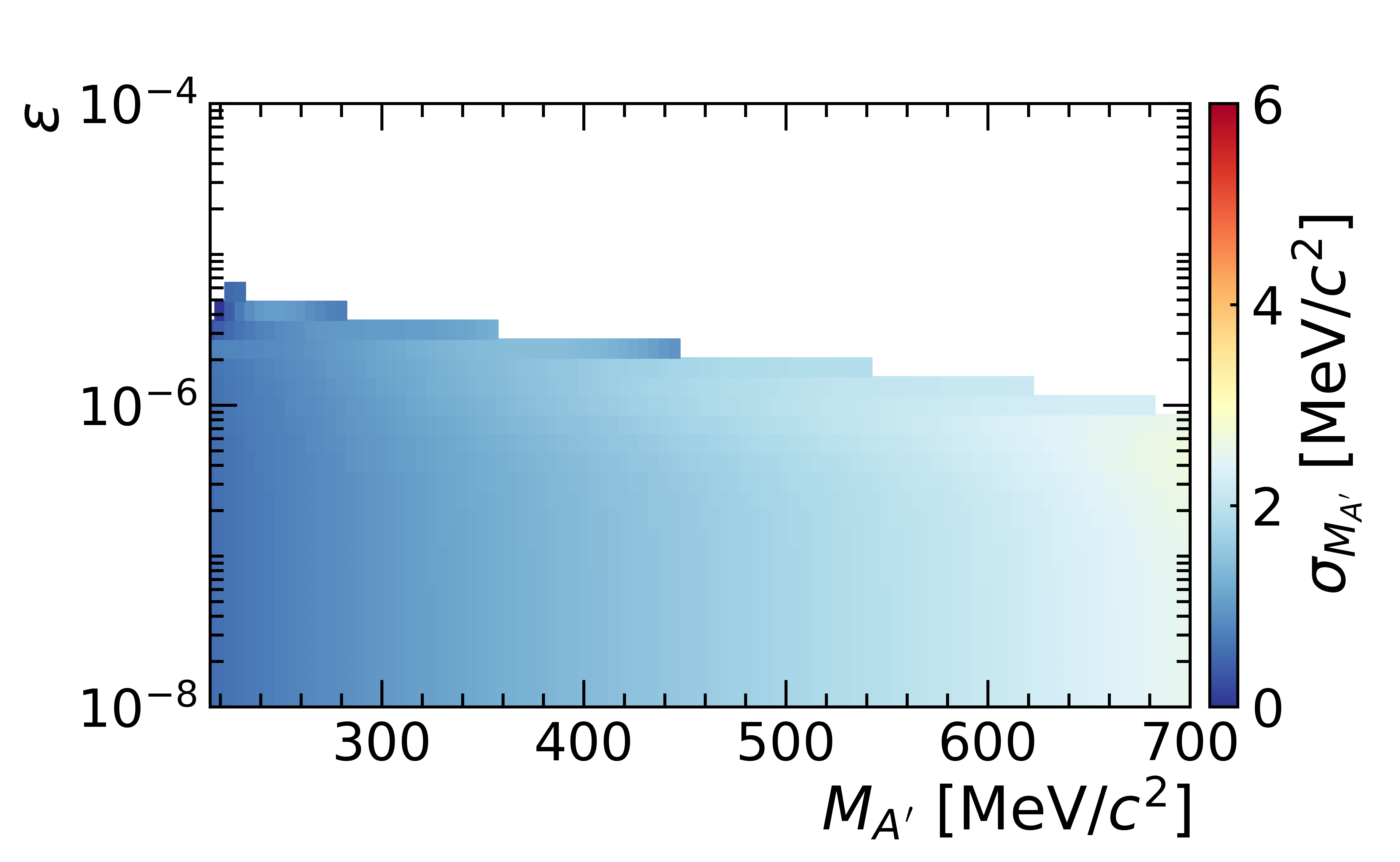}
\caption{\label{fig:sigmaSignal} Mass resolution as a function of the $A^\prime$ mass and coupling constant. Left (right) panel refers to bremsstrahlung (meson-decay) production.}
\end{figure}

\begin{figure}[t]
\centering
\includegraphics[width=.49\linewidth]{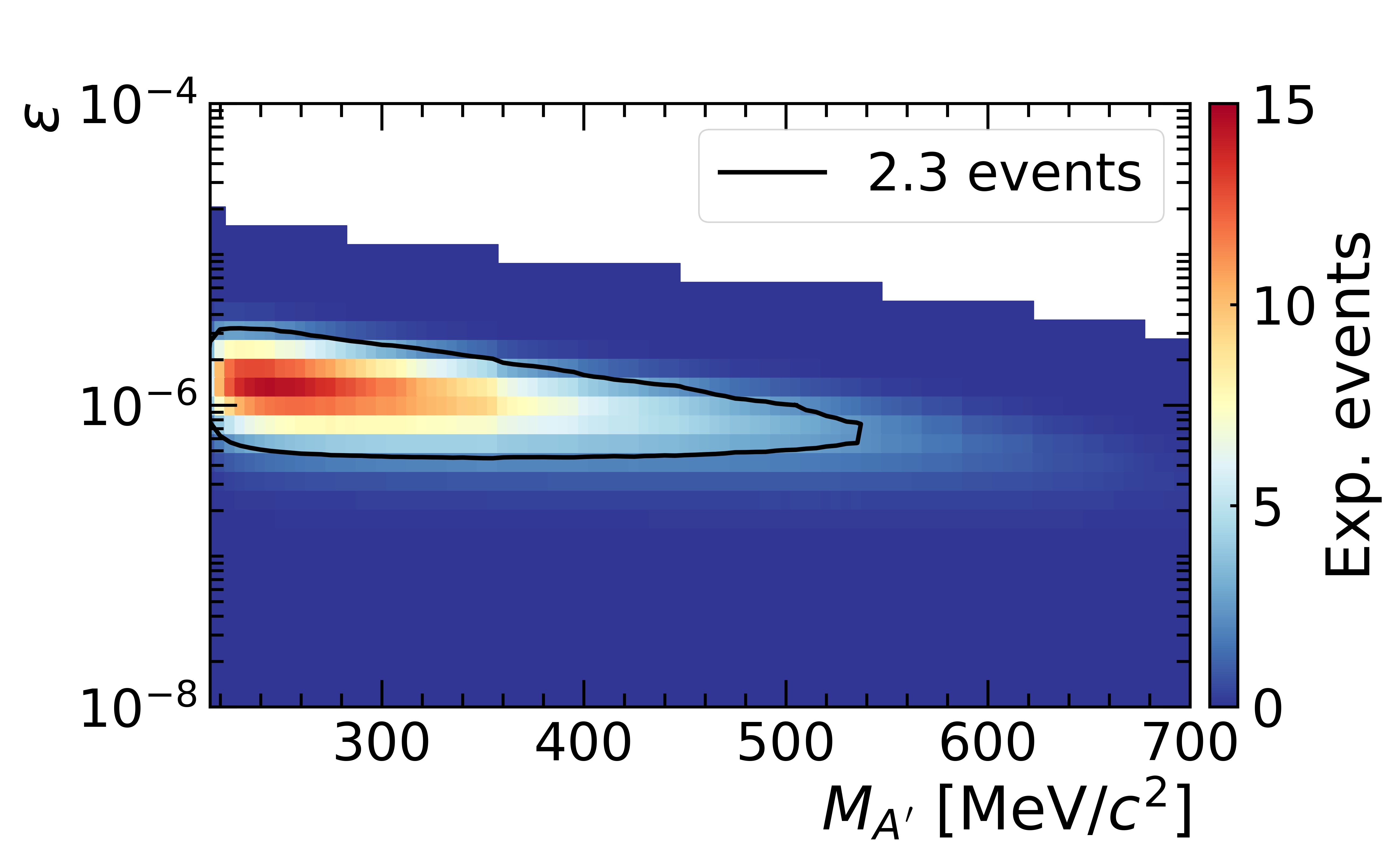}
\includegraphics[width=.49\linewidth]{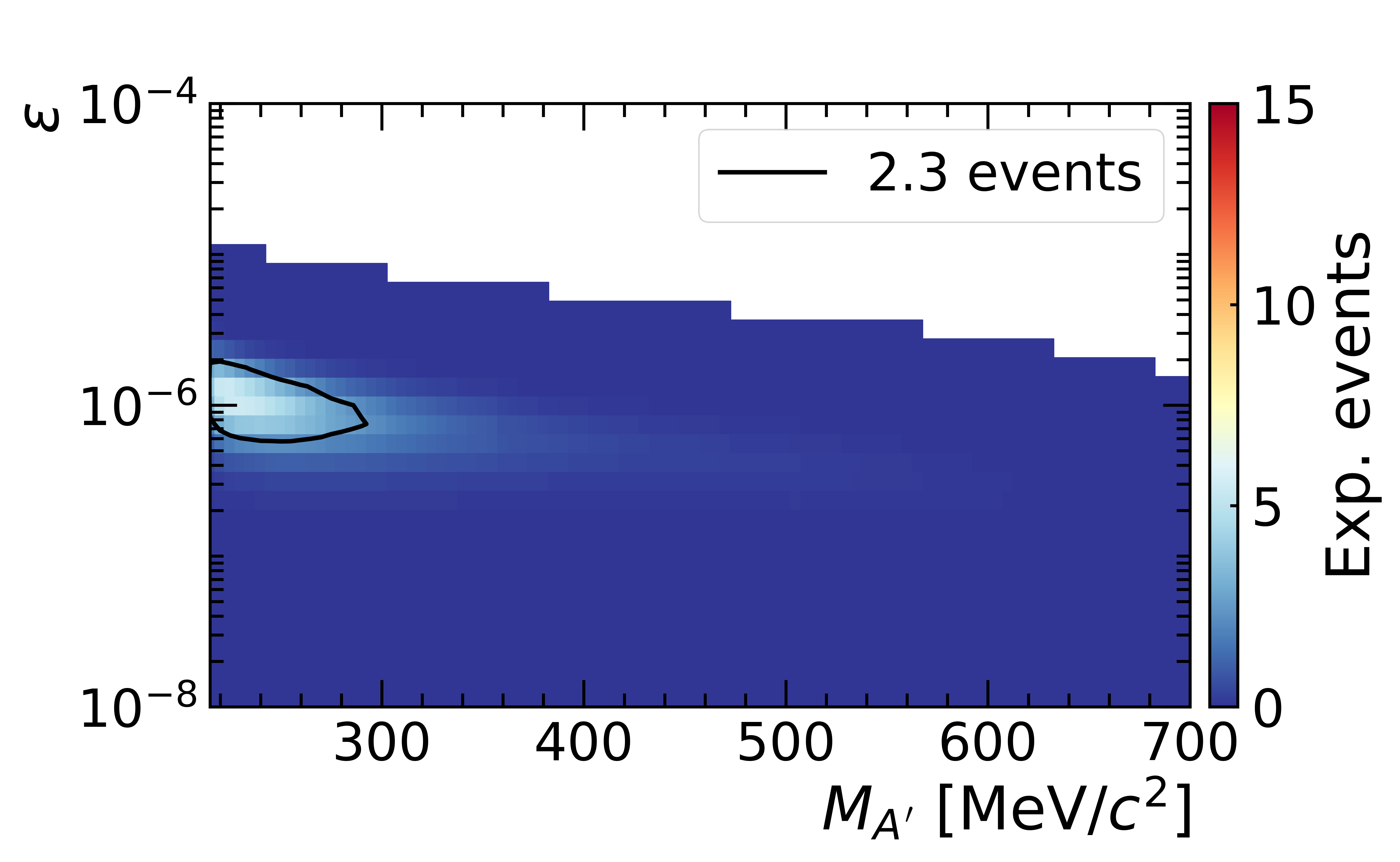}
\caption{\label{fig:signalYield} Expected number of events for the $A^\prime$ decay to $\mu^+\mu^-$ as a function of the $A^\prime$ mass and coupling constant. Left (right) panel refers to bremsstrahlung (meson-decay) production. The black contours correspond to 2.3 events.}
\end{figure}

Bremsstrahlung and meson-decay production are characterised by different $A^{\prime}$ mass  resolution, 
$\sigma_{M_{A^\prime}}$, with the former larger than the latter for most of the parameter space (figure~\ref{fig:sigmaSignal}).
The expected signal yields for the two production mechanisms are shown as functions of the $A^\prime$ coupling constant and mass in figure~\ref{fig:signalYield}.
The bremsstrahlung process dominates for most of the parameter space. Both the $A^\prime$ mass resolution and the expected signal yield are parameterised as two-dimensional 
functions of the dark photon coupling constant and mass.

\section{Results}
\begin{figure}[t]
\centering
\includegraphics[width=.5\linewidth]{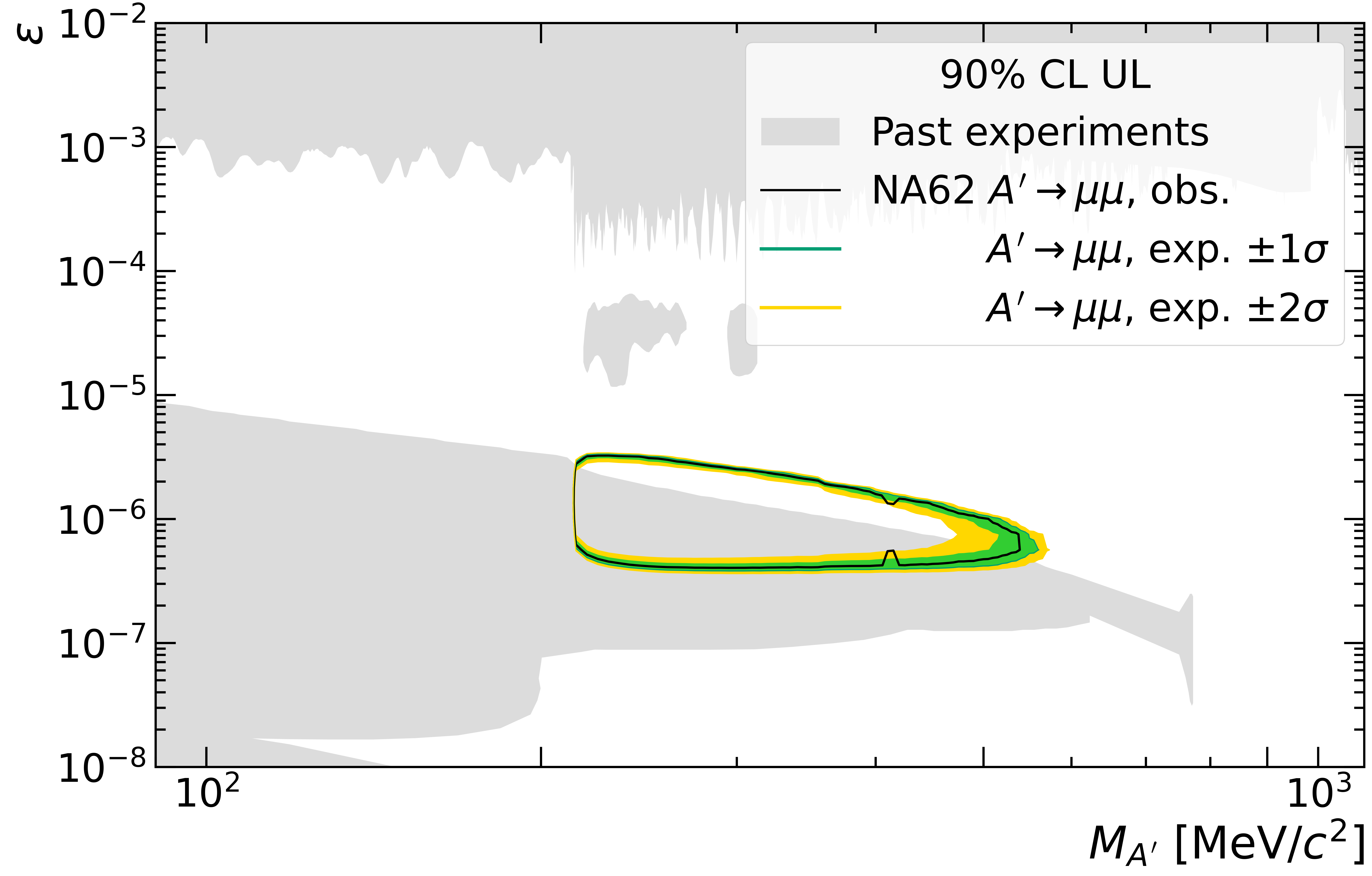}
\caption{\label{fig:obsUL} The region of the parameter space within the solid line is excluded at 90\% CL. The colour-filled area represents the expected uncertainty in the exclusion contour in the absence of a signal: green (yellow) corresponds to a statistical coverage of 68\% (95\%).}
\end{figure}

After unmasking, no events were observed in the validation region. 
The probability of a non-zero observation is  
15\%. 
After unmasking, one event was observed in the signal region. In the absence of a dark photon signal, the probability of a non-zero observation is 1.6\%.
The two-track invariant mass of the observed event is 411~MeV$/c^2$. A limit on the number of signal event counts is obtained using Poisson statistics with
negligible background, accounting for the uncertainty on the number of protons
on target (POT) using a Bayesian nuisance parameter. The corresponding observed 90\% CL upper limit is represented by the region enclosed within the black contour in figure~\ref{fig:obsUL}.  In the same figure, the colour-filled area represents the expected  uncertainty in the exclusion contour in the absence of an $A^\prime$ signal with a one-sigma (green) and two-sigma (yellow) statistical coverage.

\begin{figure}[h]
\centering
\includegraphics[width=.48\linewidth]{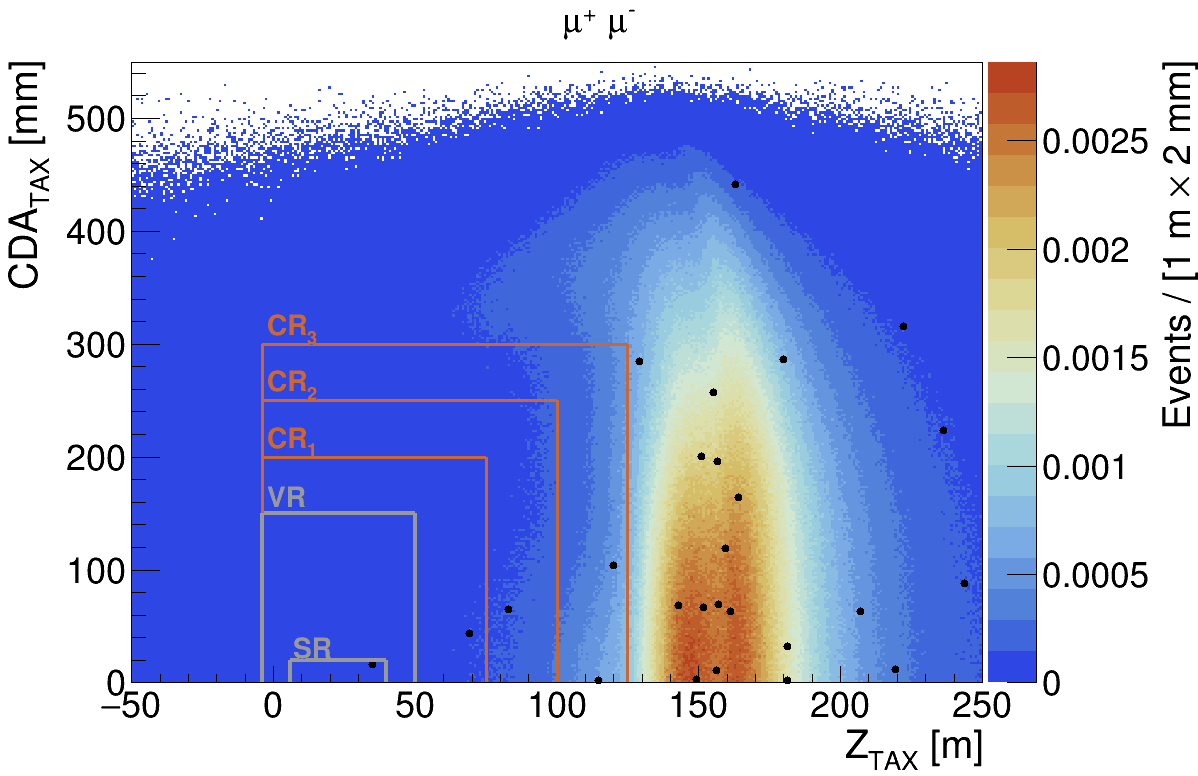}
\includegraphics[width=.48\linewidth]{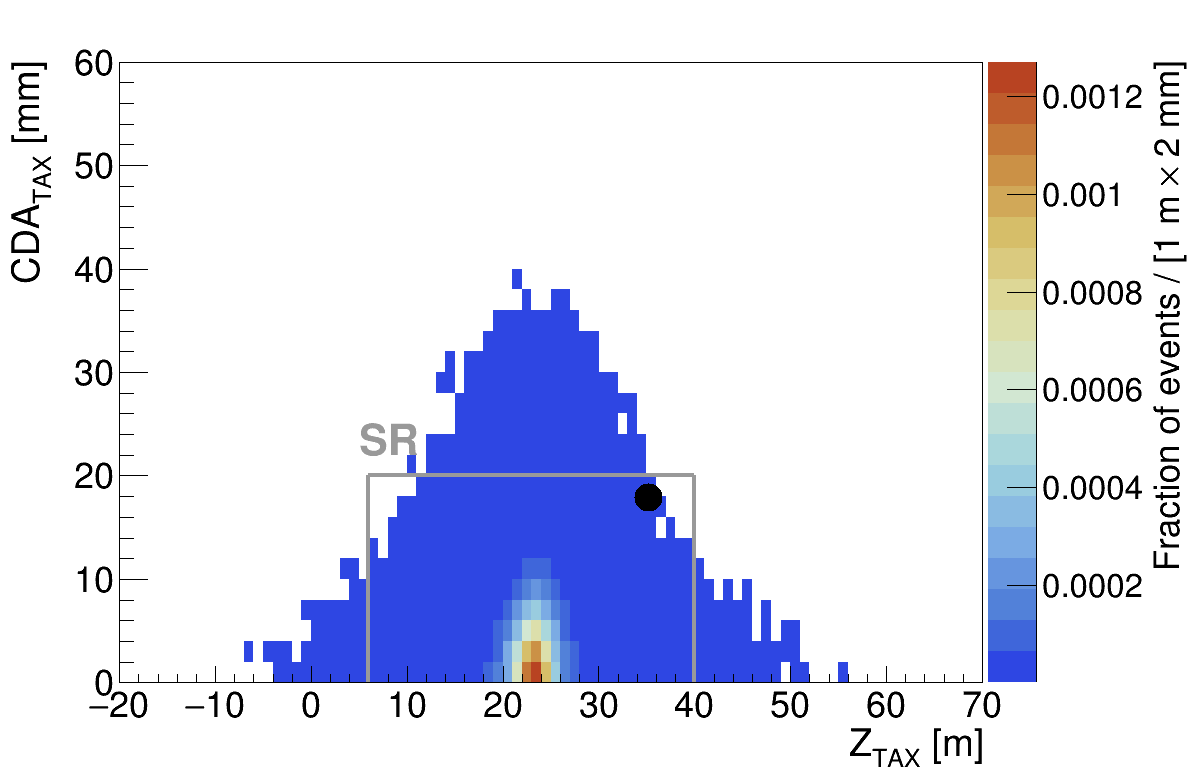}
\caption{\label{fig:SREvent} Distributions of $\mathrm{CDA}_\mathrm{TAX}$ vs $\mathrm{Z}_\mathrm{TAX}$. Left: data (dots) and expected background (colour-scale plot). Right: data (dots) and expected signal density (colour scale). Bins of $2~\mathrm{mm}\times1~\mathrm{m}$ size are used for the colour scale.}
\end{figure}

The single observed event would correspond to a 2.4 $\sigma$ global significance. 
The event observed could be interpreted as combinatorial background, since the time difference between the two tracks is $1.69$~ns, 
which is two standard deviations away from the mean for signal events.
In the $(\mathrm{Z}_\mathrm{TAX},\mathrm{CDA}_\mathrm{TAX})$ plane, 
the event observed is close to the border of the SR, consistent with the extreme tails of the expected signal (figure~\ref{fig:SREvent}).
Note that the distribution of the expected signal within the SR is not used to determine the statistical significance.

\begin{figure}[h]
\centering
\includegraphics[width=.95\linewidth]{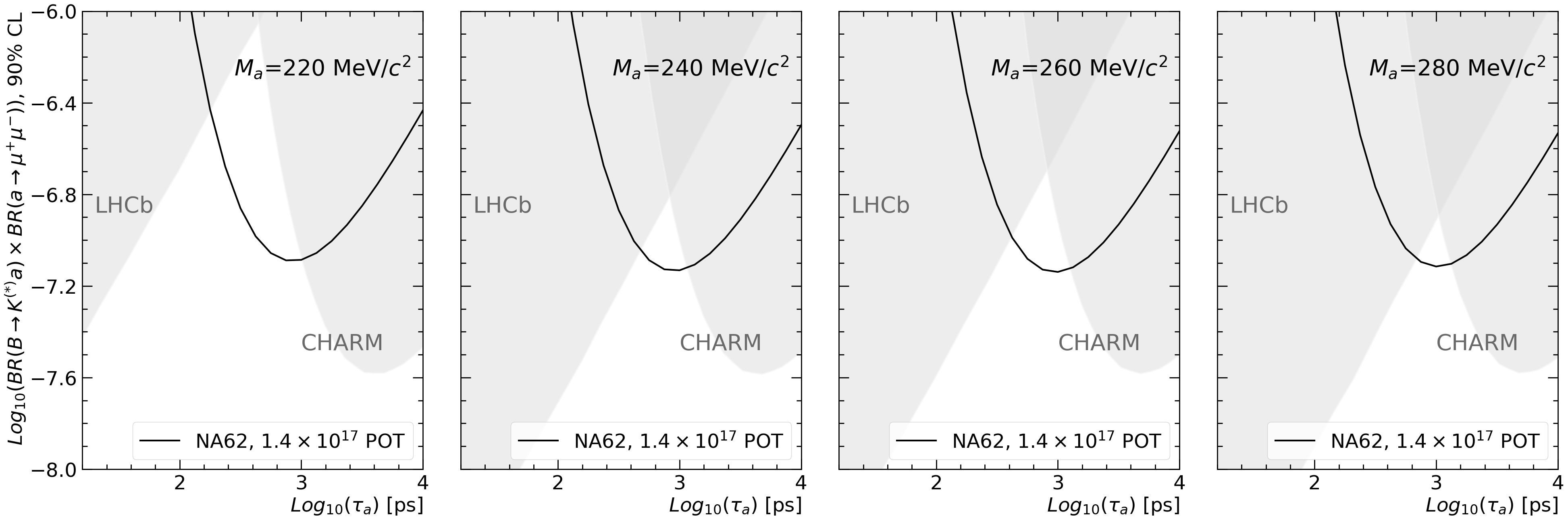}
\caption{\label{fig:alpobsUL} Search for an axion-like particle $a$ produced from decay of $B$ mesons. Four values of the ALP mass are considered. The region of the parameter space above the black line is excluded at 90\% CL. The excluded regions from LHCb~\cite{LHCb} and CHARM~\cite{CHARM} measurements 
are superimposed as grey-filled areas~\cite{ALPINIST}. 
}
\end{figure}

The results are also interpreted in terms of the emission of axion-like particles. In a model-independent approach~\cite{Dobrich:2018jyi}, the ALP lifetime $\tau_a$, the ALP mass $M_a$, and the product of the branching ratios of eq.~(\ref{eq:alpBRproduct}) are free parameters. The NA62 result is shown in figure~\ref{fig:alpobsUL} for selected values of $M_a$. For ALP masses below 280~MeV$/c^2$, the NA62 result 
extends 
the exclusion limits from previous experiments (LHCb~\cite{LHCb}, CHARM~\cite{CHARM}). 

\clearpage
\section{Conclusions}
The search for production and decay of dark photons to a di-muon final state is
the first result obtained using NA62 data taken in beam-dump mode. A counting experiment is performed through a cut-based, blind analysis of a data sample equivalent to $1.4\times10^{17}$ dumped protons. One event is found, with a possible interpretation as a combinatorial background. No evidence of a dark photon signal is established.  A region of the dark photon parameter space (coupling constant, mass) is excluded at 90\% CL, extending the limits of previous experiments in the mass range 215--550~MeV$/c^2$ for coupling constants of the order of $10^{-6}$. In addition, the result is interpreted in terms of the emission of axion-like particles in a model-independent approach. The result is found to improve on previous limits for masses below 280~MeV$/c^2$. 
%
\section*{Acknowledgements} 
It is a pleasure to express our appreciation to the staff of the CERN laboratory and the technical
staff of the participating laboratories and universities for their efforts in the operation of the
experiment and data processing.

The cost of the experiment and its auxiliary systems was supported by the funding agencies of 
the Collaboration Institutes. We are particularly indebted to: 
F.R.S.-FNRS (Fonds de la Recherche Scientifique - FNRS), under Grants No. 4.4512.10, 1.B.258.20, Belgium;
CECI (Consortium des Equipements de Calcul Intensif), funded by the Fonds de la Recherche Scientifique de Belgique (F.R.S.-FNRS) under Grant No. 2.5020.11 and by the Walloon Region, Belgium;
NSERC (Natural Sciences and Engineering Research Council), funding SAPPJ-2018-0017,  Canada;
MEYS (Ministry of Education, Youth and Sports) funding LM 2018104, Czech Republic;
BMBF (Bundesministerium f\"{u}r Bildung und Forschung) contracts 05H12UM5, 05H15UMCNA and 05H18UMCNA, Germany;
INFN  (Istituto Nazionale di Fisica Nucleare),  Italy;
MIUR (Ministero dell'Istruzione, dell'Universit\`a e della Ricerca),  Italy;
CONACyT  (Consejo Nacional de Ciencia y Tecnolog\'{i}a),  Mexico;
IFA (Institute of Atomic Physics) Romanian 
CERN-RO Nr. 10/10.03.2020
and Nucleus Programme PN 19 06 01 04,  Romania;
MESRS  (Ministry of Education, Science, Research and Sport), Slovakia; 
CERN (European Organization for Nuclear Research), Switzerland; 
STFC (Science and Technology Facilities Council), United Kingdom;
NSF (National Science Foundation) Award Numbers 1506088 and 1806430,  U.S.A.;
ERC (European Research Council)  ``UniversaLepto'' advanced grant 268062, ``KaonLepton'' starting grant 336581, Europe.

Individuals have received support from:
Charles University Research Center (UNCE/SCI/013), Czech Republic;
Ministero dell'Istruzione, dell'Universit\`a e della Ricerca (MIUR  ``Futuro in ricerca 2012''  grant RBFR12JF2Z, Project GAP), Italy;
the Royal Society  (grants UF100308, UF0758946), United Kingdom;
STFC (Rutherford fellowships ST/J00412X/1, ST/M005798/1), United Kingdom;
ERC (grants 268062,  336581 and  starting grant 802836 ``AxScale'');
EU Horizon 2020 (Marie Sk\l{}odowska-Curie grants 701386, 754496, 842407, 893101, 101023808).
\clearpage 

\clearpage

\newcommand{\orcimg}{\raisebox{-0.3\height}{\includegraphics[height=\fontcharht\font`A]{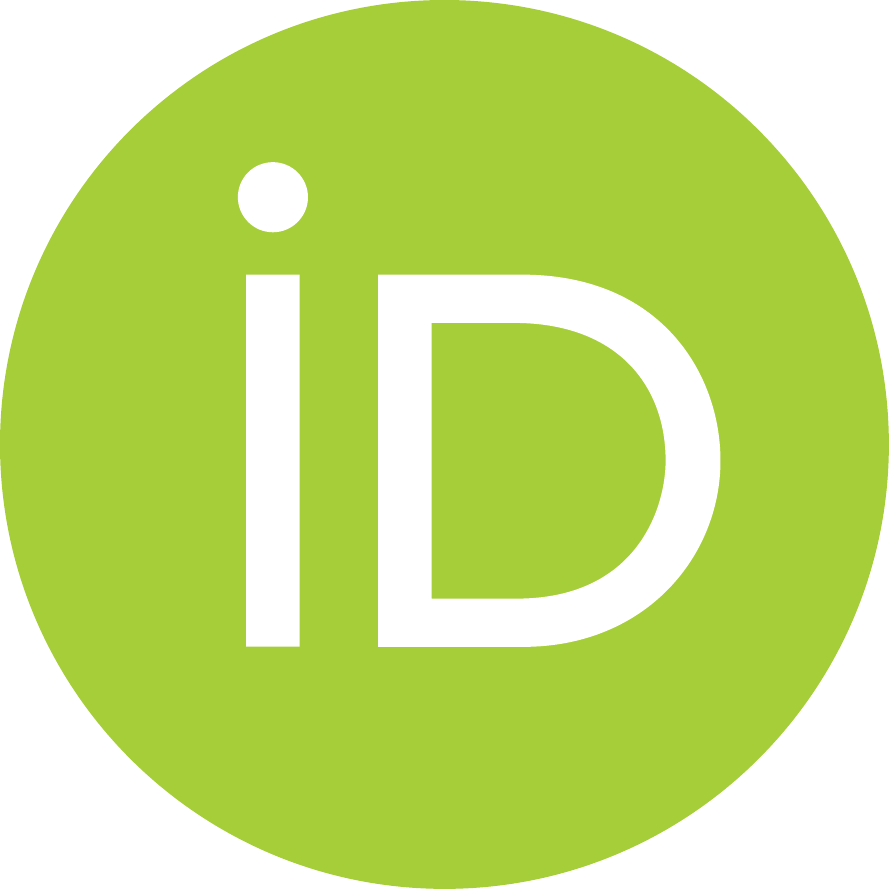}}}
\newcommand{\orcid}[1]{\href{https://orcid.org/#1}{\orcimg}}

\centerline{\bf The NA62 Collaboration} 
\vspace{1.5cm}
%
%

\begin{raggedright}
\noindent
{\bf Universit\'e Catholique de Louvain, Louvain-La-Neuve, Belgium}\\
 E.~Cortina Gil\orcid{0000-0001-9627-699X},
 J.~Jerhot\orcid{0000-0002-3236-1471},
 A.~Kleimenova$\,${\footnotemark[1]}\orcid{0000-0002-9129-4985},
 N.~Lurkin\orcid{0000-0002-9440-5927},
 M.~Zamkovsky$\,${\footnotemark[2]}\orcid{0000-0002-5067-4789}
\vspace{0.5cm}

{\bf TRIUMF, Vancouver, British Columbia, Canada}\\
 T.~Numao\orcid{0000-0001-5232-6190},
 B.~Velghe\orcid{0000-0002-0797-8381},
 V. W. S.~Wong\orcid{0000-0001-5975-8164}
\vspace{0.5cm}

{\bf University of British Columbia, Vancouver, British Columbia, Canada}\\
 D.~Bryman$\,${\footnotemark[3]}\orcid{0000-0002-9691-0775}
\vspace{0.5cm}

{\bf Charles University, Prague, Czech Republic}\\
 Z.~Hives\orcid{0000-0002-5025-993X},
 T.~Husek\orcid{0000-0002-7208-9150},
 K.~Kampf\orcid{0000-0003-1096-667X},
 M.~Koval\orcid{0000-0002-6027-317X}
\vspace{0.5cm}

{\bf Aix Marseille University, CNRS/IN2P3, CPPM, Marseille, France}\\
 B.~De Martino\orcid{0000-0003-2028-9326},
 M.~Perrin-Terrin\orcid{0000-0002-3568-1956}
\vspace{0.5cm}

{\bf Institut f\"ur Physik and PRISMA Cluster of Excellence, Universit\"at Mainz, Mainz, Germany}\\
 A.T.~Akmete\orcid{0000-0002-5580-5477},
 R.~Aliberti$\,${\footnotemark[4]}\orcid{0000-0003-3500-4012},
 L.~Di Lella\orcid{0000-0003-3697-1098},
 N.~Doble\orcid{0000-0002-0174-5608},
 L.~Peruzzo\orcid{0000-0002-4752-6160}, 
 S.~Schuchmann\orcid{0000-0002-8088-4226},
 H.~Wahl\orcid{0000-0003-0354-2465},
 R.~Wanke\orcid{0000-0002-3636-360X}
\vspace{0.5cm}

{\bf Dipartimento di Fisica e Scienze della Terra dell'Universit\`a e INFN, Sezione di Ferrara, Ferrara, Italy}\\
 P.~Dalpiaz,
 A.~Mazzolari\orcid{0000-0003-0804-6778},
 I.~Neri\orcid{0000-0002-9669-1058},
 F.~Petrucci\orcid{0000-0002-7220-6919},
 M.~Soldani\orcid{0000-0003-4902-943X}
\vspace{0.5cm}

{\bf INFN, Sezione di Ferrara, Ferrara, Italy}\\
 L.~Bandiera\orcid{0000-0002-5537-9674},
 A.~Cotta Ramusino\orcid{0000-0003-1727-2478},
 A.~Gianoli\orcid{0000-0002-2456-8667},
 M.~Romagnoni\orcid{0000-0002-2775-6903},
 A.~Sytov\orcid{0000-0001-8789-2440}
\vspace{0.5cm}

{\bf Dipartimento di Fisica e Astronomia dell'Universit\`a e INFN, Sezione di Firenze, Sesto Fiorentino, Italy}\\
 M.~Lenti\orcid{0000-0002-2765-3955},
 P.~Lo Chiatto\orcid{0000-0002-4177-557X},
 R.~Marchevski$\,${\footnotemark[1]}\orcid{0000-0003-3410-0918},
 I.~Panichi\orcid{0000-0001-7749-7914},
 G.~Ruggiero\orcid{0000-0001-6605-4739}
\vspace{0.5cm}

{\bf INFN, Sezione di Firenze, Sesto Fiorentino, Italy}\\
 A.~Bizzeti$\,${\footnotemark[5]}\orcid{0000-0001-5729-5530},
 F.~Bucci\orcid{0000-0003-1726-3838}
\vspace{0.5cm}

{\bf Laboratori Nazionali di Frascati, Frascati, Italy}\\
 A.~Antonelli\orcid{0000-0001-7671-7890},
 V.~Kozhuharov$\,${\footnotemark[6]}\orcid{0000-0002-0669-7799},
 G.~Lanfranchi\orcid{0000-0002-9467-8001},
 S.~Martellotti\orcid{0000-0002-4363-7816},
 M.~Moulson\orcid{0000-0002-3951-4389}, 
 T.~Spadaro$\,$\renewcommand{\thefootnote}{\fnsymbol{footnote}}\footnotemark[1]\renewcommand{\thefootnote}{\arabic{footnote}}\orcid{0000-0002-7101-2389},
 G.~Tinti\orcid{0000-0003-1364-844X}
\vspace{0.5cm}

{\bf Dipartimento di Fisica ``Ettore Pancini'' e INFN, Sezione di Napoli, Napoli, Italy}\\
 F.~Ambrosino\orcid{0000-0001-5577-1820},
 M.~D'Errico\orcid{0000-0001-5326-1106},
 R.~Fiorenza$\,${\footnotemark[7]}\orcid{0000-0003-4965-7073},
 R.~Giordano\orcid{0000-0002-5496-7247},
 P.~Massarotti\orcid{0000-0002-9335-9690}, 
 M.~Mirra\orcid{0000-0002-1190-2961},
 M.~Napolitano\orcid{0000-0003-1074-9552},
 I.~Rosa\orcid{0009-0002-7564-182},
 G.~Saracino\orcid{0000-0002-0714-5777}
\vspace{0.5cm}

{\bf Dipartimento di Fisica e Geologia dell'Universit\`a e INFN, Sezione di Perugia, Perugia, Italy}\\
 G.~Anzivino\orcid{0000-0002-5967-0952}
\vspace{0.5cm}

{\bf INFN, Sezione di Perugia, Perugia, Italy}\\
 F.~Brizioli$\,${\footnotemark[2]}\orcid{0000-0002-2047-441X},
 P.~Cenci\orcid{0000-0001-6149-2676},
 V.~Duk\orcid{0000-0001-6440-0087},
 R.~Lollini\orcid{0000-0003-3898-7464},
 P.~Lubrano\orcid{0000-0003-0221-4806}, 
 M.~Pepe\orcid{0000-0001-5624-4010},
 M.~Piccini\orcid{0000-0001-8659-4409}
\vspace{0.5cm}

{\bf Dipartimento di Fisica dell'Universit\`a e INFN, Sezione di Pisa, Pisa, Italy}\\
 F.~Costantini\orcid{0000-0002-2974-0067},
 M.~Giorgi\orcid{0000-0001-9571-6260},
 S.~Giudici\orcid{0000-0003-3423-7981},
 G.~Lamanna\orcid{0000-0001-7452-8498},
 E.~Lari\orcid{0000-0003-3303-0524}, 
 E.~Pedreschi\orcid{0000-0001-7631-3933},
 J.~Pinzino\orcid{0000-0002-7418-0636},
 M.~Sozzi\orcid{0000-0002-2923-1465}
\vspace{0.5cm}

{\bf INFN, Sezione di Pisa, Pisa, Italy}\\
 R.~Fantechi\orcid{0000-0002-6243-5726},
 F.~Spinella\orcid{0000-0002-9607-7920}
\vspace{0.5cm}

{\bf Scuola Normale Superiore e INFN, Sezione di Pisa, Pisa, Italy}\\
 I.~Mannelli\orcid{0000-0003-0445-7422}
\vspace{0.5cm}

{\bf Dipartimento di Fisica, Sapienza Universit\`a di Roma e INFN, Sezione di Roma I, Roma, Italy}\\
 M.~Raggi\orcid{0000-0002-7448-9481}
\vspace{0.5cm}

{\bf INFN, Sezione di Roma I, Roma, Italy}\\
 A.~Biagioni\orcid{0000-0001-5820-1209},
 P.~Cretaro\orcid{0000-0002-2229-149X},
 O.~Frezza\orcid{0000-0001-8277-1877},
 A.~Lonardo\orcid{0000-0002-5909-6508},
 M.~Turisini\orcid{0000-0002-5422-1891},
 P.~Vicini\orcid{0000-0002-4379-4563}
\vspace{0.5cm}

{\bf INFN, Sezione di Roma Tor Vergata, Roma, Italy}\\
 R.~Ammendola\orcid{0000-0003-4501-3289},
 V.~Bonaiuto$\,${\footnotemark[8]}\orcid{0000-0002-2328-4793},
 A.~Fucci,
 A.~Salamon\orcid{0000-0002-8438-8983},
 F.~Sargeni$\,${\footnotemark[9]}\orcid{0000-0002-0131-236X}
\vspace{0.5cm}

{\bf Dipartimento di Fisica dell'Universit\`a e INFN, Sezione di Torino, Torino, Italy}\\
 R.~Arcidiacono$\,${\footnotemark[10]}\orcid{0000-0001-5904-142X},
 B.~Bloch-Devaux\orcid{0000-0002-2463-1232},
 E.~Menichetti\orcid{0000-0001-7143-8200},
 E.~Migliore\orcid{0000-0002-2271-5192}
\vspace{0.5cm}

{\bf INFN, Sezione di Torino, Torino, Italy}\\
 C.~Biino\orcid{0000-0002-1397-7246},
 A.~Filippi\orcid{0000-0003-4715-8748},
 F.~Marchetto\orcid{0000-0002-5623-8494},
 D.~Soldi\orcid{0000-0001-9059-4831}
\vspace{0.5cm}

{\bf Instituto de F\'isica, Universidad Aut\'onoma de San Luis Potos\'i, San Luis Potos\'i, Mexico}\\
 A.~Briano Olvera\orcid{0000-0001-6121-3905},
 J.~Engelfried\orcid{0000-0001-5478-0602},
 N.~Estrada-Tristan$\,${\footnotemark[11]}\orcid{0000-0003-2977-9380},
 R.~Piandani\orcid{0000-0003-2226-8924},
 M. A.~Reyes Santos$\,${\footnotemark[11]}\orcid{0000-0003-1347-2579}
\vspace{0.5cm}

{\bf Horia Hulubei National Institute for R\&D in Physics and Nuclear Engineering, Bucharest-Magurele, Romania}\\
 P.~Boboc\orcid{0000-0001-5532-4887},
 A. M.~Bragadireanu,
 S. A.~Ghinescu\orcid{0000-0003-3716-9857},
 O. E.~Hutanu
\vspace{0.5cm}

{\bf Faculty of Mathematics, Physics and Informatics, Comenius University, Bratislava, Slovakia}\\
 T.~Blazek\orcid{0000-0002-2645-0283},
 V.~Cerny\orcid{0000-0003-1998-3441},
 Z.~Kucerova$\,${\footnotemark[2]}\orcid{0000-0001-8906-3902},
 R.~Volpe$\,${\footnotemark[12]}\orcid{0000-0003-1782-2978}
\vspace{0.5cm}

{\bf CERN, European Organization for Nuclear Research, Geneva, Switzerland}\\
 J.~Bernhard\orcid{0000-0001-9256-971X},
 L.~Bician$\,${\footnotemark[13]}\orcid{0000-0001-9318-0116},
 M.~Boretto\orcid{0000-0001-5012-4480},
 A.~Ceccucci\orcid{0000-0002-9506-866X},
 M.~Ceoletta\orcid{0000-0002-2532-0217}, 
 M.~Corvino\orcid{0000-0002-2401-412X},
 H.~Danielsson\orcid{0000-0002-1016-5576},
 F.~Duval,
 B.~D\"obrich$\,$\renewcommand{\thefootnote}{\fnsymbol{footnote}}\footnotemark[1]\renewcommand{\thefootnote}{\arabic{footnote}}$^,$$\,${\footnotemark[14]}\orcid{0000-0002-6008-8601},
 L.~Federici\orcid{0000-0002-3401-9522}, 
 E.~Gamberini\orcid{0000-0002-6040-4985},
 R.~Guida,
 E.~B.~Holzer\orcid{0000-0003-2622-6844},
 B.~Jenninger,
 G.~Lehmann Miotto\orcid{0000-0001-9045-7853}, 
 P.~Lichard\orcid{0000-0003-2223-9373},
 K.~Massri\orcid{0000-0001-7533-6295},
 E.~Minucci$\,$\renewcommand{\thefootnote}{\fnsymbol{footnote}}\footnotemark[1]\renewcommand{\thefootnote}{\arabic{footnote}}$^,$$\,${\footnotemark[15]}\orcid{0000-0002-3972-6824},
 M.~Noy,
 V.~Ryjov, 
 J.~Swallow\orcid{0000-0002-1521-0911}
\vspace{0.5cm}
\newpage
{\bf School of Physics and Astronomy, University of Birmingham, Birmingham, United Kingdom}\\
 J. R.~Fry\orcid{0000-0002-3680-361X},
 F.~Gonnella\orcid{0000-0003-0885-1654},
 E.~Goudzovski\orcid{0000-0001-9398-4237},
 J.~Henshaw\orcid{0000-0001-7059-421X},
 C.~Kenworthy\orcid{0009-0002-8815-0048}, 
 C.~Lazzeroni\orcid{0000-0003-4074-4787},
 C.~Parkinson\orcid{0000-0003-0344-7361},
 A.~Romano\orcid{0000-0003-1779-9122},
 J.~Sanders\orcid{0000-0003-1014-094X},
 A.~Sergi$\,${\footnotemark[16]}\orcid{0000-0001-9495-6115}, 
 A.~Shaikhiev$\,${\footnotemark[17]}\orcid{0000-0003-2921-8743},
 A.~Tomczak\orcid{0000-0001-5635-3567}
\vspace{0.5cm}

{\bf School of Physics, University of Bristol, Bristol, United Kingdom}\\
 H.~Heath\orcid{0000-0001-6576-9740}
\vspace{0.5cm}

{\bf School of Physics and Astronomy, University of Glasgow, Glasgow, United Kingdom}\\
 D.~Britton\orcid{0000-0001-9998-4342},
 A.~Norton\orcid{0000-0001-5959-5879},
 D.~Protopopescu\orcid{0000-0002-3964-3930}
\vspace{0.5cm}

{\bf Faculty of Science and Technology, University of Lancaster, Lancaster, United Kingdom}\\
 J. B.~Dainton,
 L.~Gatignon\orcid{0000-0001-6439-2945},
 R. W. L.~Jones\orcid{0000-0002-6427-3513}
\vspace{0.5cm}

{\bf Physics and Astronomy Department, George Mason University, Fairfax, Virginia, USA}\\
 P.~Cooper,
 D.~Coward$\,${\footnotemark[18]}\orcid{0000-0001-7588-1779},
 P.~Rubin\orcid{0000-0001-6678-4985}
\vspace{0.5cm}

{\bf Authors affiliated with an Institute or an international laboratory covered by a cooperation agreement with CERN}\\
 A.~Baeva,
 D.~Baigarashev$\,${\footnotemark[19]}\orcid{0000-0001-6101-317X},
 D.~Emelyanov,
 T.~Enik\orcid{0000-0002-2761-9730},
 V.~Falaleev$\,${\footnotemark[12]}\orcid{0000-0003-3150-2196}, 
 S.~Fedotov,
 K.~Gorshanov\orcid{0000-0001-7912-5962},
 E.~Gushchin\orcid{0000-0001-8857-1665},
 V.~Kekelidze\orcid{0000-0001-8122-5065},
 D.~Kereibay, 
 S.~Kholodenko$\,${\footnotemark[20]}\orcid{0000-0002-0260-6570},
 A.~Khotyantsev,
 A.~Korotkova,
 Y.~Kudenko\orcid{0000-0003-3204-9426},
 V.~Kurochka, 
 V.~Kurshetsov\orcid{0000-0003-0174-7336},
 L.~Litov$\,${\footnotemark[6]}\orcid{0000-0002-8511-6883},
 D.~Madigozhin\orcid{0000-0001-8524-3455},
 A.~Mefodev,
 M.~Misheva$\,${\footnotemark[21]}, 
 N.~Molokanova,
 V.~Obraztsov\orcid{0000-0002-0994-3641},
 A.~Okhotnikov\orcid{0000-0003-1404-3522},
 I.~Polenkevich,
 Yu.~Potrebenikov\orcid{0000-0003-1437-4129}, 
 A.~Sadovskiy\orcid{0000-0002-4448-6845},
 S.~Shkarovskiy,
 V.~Sugonyaev\orcid{0000-0003-4449-9993},
 O.~Yushchenko\orcid{0000-0003-4236-5115}
\vspace{0.5cm}

\end{raggedright}
\vspace{0.5cm}
%
%

\setcounter{footnote}{0}
\newlength{\basefootnotesep}
\setlength{\basefootnotesep}{\footnotesep}

\renewcommand{\thefootnote}{\fnsymbol{footnote}}
\noindent
$^{\footnotemark[1]}${Corresponding authors:  B.~D\"obrich, E.~Minucci, T.~Spadaro, email: babette.dobrich@cern.ch, elisa.minucci@cern.ch, tommaso.spadaro@cern.ch}\\
\renewcommand{\thefootnote}{\arabic{footnote}}
$^{1}${Present address: Ecole Polytechnique F\'ed\'erale Lausanne, CH-1015 Lausanne, Switzerland} \\
$^{2}${Present address: CERN, European Organization for Nuclear Research, CH-1211 Geneva 23, Switzerland} \\
$^{3}${Also at TRIUMF, Vancouver, British Columbia, V6T 2A3, Canada} \\
$^{4}${Present address: Institut f\"ur Kernphysik and Helmholtz Institute Mainz, Universit\"at Mainz, Mainz, D-55099, Germany} \\
$^{5}${Also at Dipartimento di Scienze Fisiche, Informatiche e Matematiche, Universit\`a di Modena e Reggio Emilia, I-41125 Modena, Italy} \\
$^{6}${Also at Faculty of Physics, University of Sofia, BG-1164 Sofia, Bulgaria} \\
$^{7}${Present address: Scuola Superiore Meridionale e INFN, Sezione di Napoli, I-80138 Napoli, Italy} \\
$^{8}${Also at Department of Industrial Engineering, University of Roma Tor Vergata, I-00173 Roma, Italy} \\
$^{9}${Also at Department of Electronic Engineering, University of Roma Tor Vergata, I-00173 Roma, Italy} \\
$^{10}${Also at Universit\`a degli Studi del Piemonte Orientale, I-13100 Vercelli, Italy} \\
$^{11}${Also at Universidad de Guanajuato, 36000 Guanajuato, Mexico} \\
$^{12}${Present address: INFN, Sezione di Perugia, I-06100 Perugia, Italy} \\
$^{13}${Present address: Charles University, 116 36 Prague 1, Czech Republic} \\
$^{14}${Present address: Max-Planck-Institut f\"ur Physik (Werner-Heisenberg-Institut), M\"unchen, D-80805, Germany} \\
$^{15}${Present address: Syracuse University, Syracuse, NY 13244, USA} \\
$^{16}${Present address: Dipartimento di Fisica dell'Universit\`a e INFN, Sezione di Genova, I-16146 Genova, Italy} \\
$^{17}${Present address: Faculty of Science and Technology, University of Lancaster, Lancaster, LA1 4YW, UK} \\
$^{18}${Also at SLAC National Accelerator Laboratory, Stanford University, Menlo Park, CA 94025, USA} \\
$^{19}${Also at L.N. Gumilyov Eurasian National University, 010000 Nur-Sultan, Kazakhstan} \\
$^{20}${Present address: INFN, Sezione di Pisa, I-56100 Pisa, Italy} \\
$^{21}${Present address: Institute of Nuclear Research and Nuclear Energy of Bulgarian Academy of Science (INRNE-BAS), BG-1784 Sofia, Bulgaria} \\


\begin{thebibliography}{99}

\bibitem{Okun} L. Okun, Sov. Phys. JETP 56 (1982) 502.

\bibitem{Holdom} B. Holdom, Phys. Lett. B 166 (1986) 196.

\bibitem{GalisonManohar} {P.~Galison and A.~Manohar,
Phys. Lett. B 136 (1984) 279.}

\bibitem{Rella:2022len}
C.~Rella, B.~D\"obrich and T.~T.~Yu,
Phys. Rev. D 106 (2022) 035023.

\bibitem{Dobrich:2018jyi}
B.~D\"obrich, F.~Ertas, F.~Kahlhoefer and T.~Spadaro,
Phys. Lett. B 790 (2019) 537.


\bibitem{PBCConventionalBeam} 
L. Gatignon et al., 
CERN-PBC-REPORT-2018-002,
\href{https://cds.cern.ch/record/2650989}{Report from the Conventional Beams Working Group to the Physics beyond Collider Study and to the European Strategy for Particle Physics}.

\bibitem{Rosental} M. Rosental et al., 
Int. J. Mod. Phys. A 34 (2019) 1942026.




\bibitem{na62det}
E.~Cortina Gil et al. [The NA62 Collaboration],
JINST 12 (2017) P05025.

\bibitem{ANTI0}
H.~Danielsson et al., 
JINST 15 (2020) C07007.

\bibitem{na62trigperf}
E.~Cortina Gil et al. [The NA62 Collaboration],
arXiv:2208.00897 [hep-ex], submitted to JHEP.

\bibitem{tdaq}
R.~Ammendola et al.,
Nucl. Instrum. Meth. A 929 (2019) 1.

\bibitem{GEANT4}
J. Allison et al.,
Nucl. Instrum. Meth. A 835 (2016) 186.

\bibitem{BlumleinBrunner13}
J. Bl{\"u}mlein and J. Brunner, 
Phys. Lett. B 731 (2014) 320.

\bibitem{pythia}
T. Sjöstrand et al., 
Comput. Phys. Commun. 191 (2015) 159.

\bibitem{Dobrich:2019dxc}
B.~D\"obrich, J.~Jaeckel and T.~Spadaro,
JHEP 05 (2019) 213 
[erratum: JHEP 10 (2020) 046].

\bibitem{Darkcast} 
P.~Ilten, Y.~Soreq, M.~Williams and W.~Xue,
 JHEP 6 (2018) 004.



\bibitem{PBCSMWG} 
J.~Beacham et al.,
J. Phys. G 47 (2020) 010501.


\bibitem{JHEPMC} S.~Ghinescu, B.~D\"obrich, E.~Minucci and T.~Spadaro,
Eur. Phys. J. C 81 (2021) 767.

\bibitem{PUMAS} 
V. Niess, A. Barnoud, C. C\^{a}rloganu, E. Le M\'{e}n\'{e}deu,
Computer Physics Communications, 229 (2018) 54.







\bibitem{LHCb}  R. Aaij et al. [The LHCb Collaboration], Phys. Rev. Lett. 115 (2015) 161802; Phys. Rev. D 95 (2017) 071101.

\bibitem{CHARM}  F. Bergsma et al. [The CHARM Collaboration], Phys. Lett. B 157 (1985) 458; Phys. Lett. B, 128 (1983) 361.

\bibitem{ALPINIST} {J. Jerhot et al., \href{https://doi.org/10.5281/zenodo.7963458}{zenodo (2023) https://doi.org/10.5281/zenodo.7963458}.}



\end{thebibliography}
\end{document}